%
%


\def\Z{{\rm Z\!\!\rm Z}}	

\def\bflb{{\bf [}}		
\def\bfrb{{\bf ]}}		
\def\Aut{{\rm Aut}}		
\def\half{{1 \over 2}}		

\def\One{{\bf 1}}		
\def\ideq{\equiv}		
\def\tilde{\widetilde}		
\def\hat{\widehat}		

\def\semidirect{\mathbin
               {\hbox
               {\hskip 3pt \vrule height 5.2pt depth -.2pt width .55pt 
                \hskip-1.5pt$\times$}}}

\def\Nobreak{\par\nobreak}	
\def\mod{{\rm mod}}		
\def\card{{\rm card}}		
\def\Diff{{\rm Diff}}		
\def\isom{\simeq}		
\def\R{{\rm I\!\rm R}}		
\def\M{{\cal M}}		
\def\D{{\bf D}}			
\def\O{{\cal O}}		
\def\cO{{\cal O}}		
\def\tG{{\widetilde G}}		

\def\capG{{G_{int}}}		
\def\internals{{G_{int}}}	

\def\perms{S_N}			

\def\hG{{\hat G}}		
\def\hN{{\hat N}}		

\def\cH{{\cal H}}		
\def\cP{{\cal P}}		
\def\tR{{\widetilde R}}
\def\cC{{\rm C}\llap{\vrule height6.3pt width1pt depth-.4pt\phantom t}}
\def\cE{{\cal E}}
\def\hcS{{\hat {\cal S}}} 
\def\cS{{\cal S}}		
\def\slides{{\cal S}}		

\def\xsq{\otimes}		
\def\A{{\cal A}}
\def\B{{\cal B}}
%


\input harvmac
 



\raggedbottom
\hsize=6.5 true in
\vsize=9 true in
\voffset=0.3 true cm                

\def\singlespace{\baselineskip=12pt}  
\def\sesquispace{\baselineskip=16pt}    
\noblackbox

\parindent 35pt
\parskip 2pt
\lineskip 2pt

\input epsf

\sesquispace


\lref\wormhole{
  S.W.~Hawking and R.~Laflamme,
  ``Baby Universes and the Nonrenormalizability of Gravity''
    {\it Phys. Lett.} {\bf 209B}~:~39 (1988);
  Sidney Coleman,
  ``Black Holes as Red Herrings'',
   {\it Nuclear  Physics} {\bf B 307}: 867-882 (1988);
 Steven B. Giddings and Andrew Strominger, 
 ``Loss of Incoherence and Determination of Coupling Constants
   in Quantum Gravity'', 
  {\it Nuclear Physics} {\bf B 307}: 854-866 (1988) }
          
\lref\BalPrivate{
	A.P.~Balachandran (private communication)}
 
\lref\ArvindRafael{
	Arvind Borde and Rafael~D.~Sorkin,  
  	``Causal Cobordism: Topology Change Without Causal Anomalies'',
    	(in preparation)}
 
\lref\clifford{
	A.H. Clifford, 
	``Representations Induced in an Invariant Subgroup'', 
	{\it Annals of Mathematics} {\bf 38:} 533-550 (1937)}
 
\lref\thesis{
	Sumati Surya, Ph.D. Thesis (in preparation)}
 
\lref\AmalfiBal{
  	A.P. Balachandran, 
 	``Classical Topology and Quantum Phases: Quantum Mechanics''
   	in Filippo, S. de, Marinaro, M., Marmo, G., (eds.), 
   	{\it Geometrical and Algebraic Aspects of Nonlinear Field Theories}
   	(Proceedings of the conference of the same name, 
      	held Amalfi, Italy, May 1988), 1-28,
   	(Elsevier, Amsterdam, 1989)}
 
\lref\amalfi{
	 R.D.~Sorkin, 
 	``Classical Topology and Quantum Phases: Quantum Geons'',
   	in Filippo, S. de, Marinaro, M., Marmo, G., (eds.), 
   	{\it Geometrical and Algebraic Aspects of Nonlinear Field Theories}
   	(Proceedings of the conference of the same name, 
      	held Amalfi, Italy, May 1988), 201-218, 
   	(Elsevier, Amsterdam, 1989)  }
 
\lref\rds{ 
   	R.D.~Sorkin, 
  	``Introduction to Topological Geons'',
   	in P.G. Bergmann and V. de Sabbata (eds.),
   	{\it Topological Properties and Global Structure of Space-Time}, 
   	(Proceedings of the conference of the same name, 
      	held Erice, Italy, May 1985), 249-270,
   	(Plenum Press, 1986) }	
 
\lref\isham{ 
   	C.J. Isham,
  	``Topological and Global Aspects of Quantum Theory'',  
    	in  B.S. DeWitt and R.Stora (eds.),
   	{\it Relativit\'e, Groupes et Topologie II''}, 
    	(Proceedings of the conference of the same name,
	held LesHouches, France, 1983), 1059-1290,
    	(North-Holland, 1984) } 	
 
\lref\balrds{
   	Ch.~Aneziris, A.P.~Balachandran, M.~Bourdeau, S.~Jo, T.R.~Ramadas 
   	and R.D.~Sorkin, 
 	``Aspects of Spin and Statistics in Generally Covariant Theories'', 
   	{\it Int. J. Mod. Phys.} {\bf A4:} 5459-5510 (1989) }
 
\lref\fsor{
	{ J.L.~Friedman and R.D.~Sorkin, 
 	``Spin-1/2 from Gravity'',
  	{\it Phys. Rev. Lett.} {\bf 44:} 1100-1103 (1980); 
	and {\bf 45}:148 (1980)}
	{ J.L.~Friedman and R.D.~Sorkin, 
	``Half-Integral Spin from Quantum Gravity'', 
 	{\it Gen. Rel. Grav.} {\bf 14}:615-620 (1982)}}
 
\lref\fwit{
	{J.L. Friedman and D.M. Witt, 
 	``Internal Symmetry Groups of Quantum Geons'', 
	{\it Phys. Lett.} {\bf 120B}: 324-328 (1983)}
 	{D.M. Witt, 
	`` Symmetry Groups of State Vectors in Canonical Quantum
	Gravity'',  
	{\it J. Math. Phys.} {\bf 27}: 573-592 (1986)}}

\lref\hriks{ 
	H.~Hendriks, 
	`` Applications De La Th\'eorie D'Obstruction En Dimension 3'',  
	{\it Bull. Soc. Math. France, Memoire},{\bf 53}: 81 (1977).
        See Section 4.3.}
 
\lref\mcmil{
	D. McCullough and A. Miller, 
	``Homeomorphisms of 3-Manifolds with Compressible
	Boundary'' 
	{\it Memoires of the American Math. Soc.}, {\bf Vol. 61}, 
	{\bf No. 344} (1986) (See Chapter 5)}
 
\lref\mcc{
	D. McCullough, 
	`` Topological and Algebraic Automorphisms of Three
	Manifolds'',  in
	R.A. Piccinini (ed.), 
	{\it Groups of Self Equivalences and Related Topics}, 102-113,
	(Springer Lecture Notes in Mathematics, 1425)
	(Springer Verlag, 1988)}

\lref\hrksmc{H.~Hendriks and D. McCullough, 
	``On the Diffeomorphism Group of a Reducible Three Manifold'',
	{\it Topology Appl.}, {\bf 26}: 25-31 (1987)}
	
\lref\rdsfay{
    H.F. Dowker and R.D. Sorkin, 
  ``A Spin Statistics Theorem for Certain  Topological Geons'', 
  $<$e-print archive: gr-qc/9609064$>$}
  
\lref\nico{
	D. Giulini, 
	`` Properties of Three Manifolds for Relativists'',
	{Freiberg-THEP-93-15} $<$e-print archive: gr-qc/9308008$>$}
 
\lref\cox{
       S.M.~Coxeter and W.O.J.~Moser,
       {\it Generators and Relations for Discrete Groups}, $4^{th}$ edition
        (Springer-Verlag, 1980)}
 
\lref\mcka{
	G.W. Mackey, 
	"Induced Representations of Groups and Quantum Mechanics'', 
         (W.A. Benjamin Inc. and Editore Boringhieri, 1968);
	G.W. Mackey, ``Theory of Unitary Group Representations'', 
	(University of Chicago Press, 1976); 
	G.W. Mackey, 
	``Unitary Group Representations in Physics, Probability and
	Number Theory'',  
	Addison-Wesley Publishing Co. Inc., 1978}
 
\lref\dwitt{
	Donald M.~Witt, 
	`` Vacuum Space-Times That Admit No Maximal Slice'', 
	{\it Phys. Rev. Lett.} {\bf 57} : 1386-1389, (1986)  }
 
\lref\balsachin{
	A.P.~Balachandran and S.M. Vaidya, 
       ``Parity Doubles in Quark Physics'',
	 {\it Phys. Rev. Lett.} {\bf 78}:13-16 (1997)
         $<$e-Print Archive: hep-ph/9606283$>$;
	``Emergent Chiral Symmetry: Parity and Time Reversal Doubles''
        (Syracuse University preprint SU-4240-653)
        $<$e-print archive: hep-th/9612053$>$}
 
\lref\balwitt{
	A.P.~Balachandran, A.~Simoni and D.M.~Witt,
 	``Molecules and how they violate P and T'', 
  	{\it Int. J. Modern Phys.} {\bf A7}:2087-2107 (1992)}

\lref\posets{
	{L.~Bombelli, J.~Lee, D.~Meyer and R.D.~Sorkin, 
	``Spacetime as a causal set'', 
  	{\it Phys. Rev. Lett.} {\bf 59}:521 (1987)};
	{L.~Bombelli, 
	{\it Space-time as a Causal Set},  
  	Ph.D. dissertation (Syracuse University, 1987)};
	{D.A.~Meyer,
 	{\it The Dimension of Causal Sets},
	 Ph.D. dissertation ( M.I.T., 1988)};
	{G.~Brightwell and  R.~Gregory,
	``The Structure of Random Discrete Spacetime'',
    	{\it Phys. Rev. Lett.} {\bf 66}:260-263 (1991)};
	{A.R.~Daughton,
	``The Recovery of Locality for Causal Sets and Related Topics'',
  	Ph.D. dissertation (Syracuse University, 1993)};
	{J.~Myrheim, ``Statistical geometry,'' 
	CERN preprint TH-2538 (1978)};
	{G.~'t~Hooft, 
  	``Quantum gravity: a fundamental problem and some radical ideas,'', 
     	in M. Levy, S. Deser (eds.),
	{\it Recent Developments in Gravitation }  
       	(Proceedings of the 1978 Cargese Summer Institute),  
        (Plenum, 1979)}
	}

\lref\unimod{
	R.D.~Sorkin, 
	``On the Role of Time in the Sum-over-histories Framework for
	Gravity'', 
    	paper presented to the conference on The History of Modern
      	Gauge Theories, held Logan, Utah, July, 1987; 
        published in  
        {\it Int. J. Theor. Phys.} {\bf 33}:523-534 (1994)}
 
\lref\drexel{
  R.D.~Sorkin,
  ``Quantum Measure Theory and its Interpretation'', in
    D.H.~Feng and B-L~Hu (eds.),  
    {\it Proceedings of the Fourth Drexel Symposium on Quantum
       Nonintegrability: Quantum Classical Correspondence},
       held Philadelphia, September 8-11, 1994
    (World Scientific, in press)
    $<$e-Print Archive: gr-qc/9507057$>$}

\lref\fuksrab{
  {D.I. Fuks-Rabinovich, 
  ``On the Automorphism Groups of Free Products I''
   (``O gruppakh avtomorfismov svobodnykh proizvedenii. I'')
   {\it Mat. Sbornik}, {\bf 8}(50): 265-276 (1940)};
  {D.I. Fuks-Rabinovich, 
  ``On the Automorphism Groups of Free Products II'',
   (``O gruppakh avtomorfismov svobodnykh proizvedenii. II'')
   {\it Mat. Sbornik}, {\bf 9}(51): 183-220 (1941)}}
 
\lref\jimdon{
	J.B. Hartle and D.M. Witt,
	``Gravitational Theta States and the Wavefunction of the Universe'',
	{\it Phys. Rev. D}, {\bf 37}: 2833-2837 (1988)}

\lref\jorma{
	Domenico Giulini and Jorma Louko,
	``No Boundary Theta Sectors in Spatially Flat  Quantum Cosmology'', 
	{\it Phys. Rev. D}, {\bf 37}: 4355-4364 (1992)}
 
\lref\rourkedesa{
	Eug\'enia C\'esar de S\'a and Colin Rourke,
	`` The Homotopy Type of Homeomorphisms of 3-Manifolds'',
	{\it Bulletin of The American Mathematical Society}, 
	{\bf Vol. 1}, {\bf No. 1}: 251-254 (1979)}
 
\lref\cecile{
	Michael G.C. Laidlaw and C\'ecile Morette DeWitt,
	``Feynman Functional Integrals for Systems of
	Indistinguishable Particles'',
	{\it Phys. Rev. D}, {\bf 3}: 1375-1378 (1971)}  
 
\lref\hempel{
	J. Hempel,
	``3 - Manifolds'',
	{\it Annals of Math. Study}, {\it No. 86},
	(Princeton University Press, Princeton, 1976)}

\lref\nicoHPA{
      Domenico Giulini,
     ``On the Configuration Space Topology in General Relativity'', 
      {\it Helv. Phys. Acta}, {\bf 68}: 86-111 (1995), see pp. 91-92.}



\centerline{\phantom{abc}}

\centerline {\bf AN ANALYSIS OF THE REPRESENTATIONS OF THE}
\centerline {\bf MAPPING CLASS GROUP OF A MULTI-GEON THREE-MANIFOLD}
\singlespace
\bigskip
\centerline{ {\bf Rafael D. Sorkin}$^a$} 
\smallskip
\centerline{\sl Departamento de Gravitaci\'on y Teor{\'\i}a de Campos}
\centerline{\sl Instituto de Ciencias Nucleares, Universidad Nacional
                Aut{\'o}noma de M\'exico} 
\centerline{\sl A.~Postal 70-543, M\'exico D.F. 04510, M\'exico} 

\centerline{and}

\centerline{\sl Department of Physics, Syracuse University, 
                Syracuse, N.Y. 13244-1130} 

\centerline{{\bf  \&}}
\smallskip
\centerline{ {\bf Sumati Surya}$^b$}
\smallskip
\centerline{\sl Department of Physics, Syracuse University, Syracuse, N.Y.,
13244-1130} 
\smallskip
\centerline{\it $^a$email:
            sorkin@xochitl.nuclecu.unam.mx, sorkin@suhep.syr.edu}
\centerline{\it $^b$email: ssurya@suhep.syr.edu}

\bigskip
 

\centerline{\bf ABSTRACT}

One knows that the distinct unitary irreducible representations (UIR's) of
the mapping class group $G$ of a 3-manifold $\M$ give rise to distinct
quantum sectors (``$\theta$-sectors'') in quantum theories of gravity based
on a product spacetime of the form $\R\times\M$.  In this paper, we study
the UIR's of $G$ in an effort to understand the physical implications of
these quantum sectors.  The mapping class group of a 3-manifold which is
the connected sum of $\R^3$ with a finite number of irreducible primes is a
semi-direct product group.  Following Mackey's theory of induced
representations, we provide an analysis of the structure of the general
finite dimensional UIR of such a group.  In the picture of quantized primes
as particles (topological geons), this general group-theoretic analysis
enables one to draw several qualitative conclusions about the geons'
behavior in different quantum sectors, without requiring an explicit
knowledge of the UIR's corresponding to the individual primes.  An
important general result is that the classification of the UIR's of the so
called particle subgroup (equivalently, the UIR's of $G$ in which the slide
diffeomorphisms are represented trivially) is reduced to the problem of
finding the UIR's of the internal diffeomorphism groups of the individual
primes.  Moreover, this reduction is entirely consistent with the geon
picture, in which the UIR of the internal group of a prime determines the
{\it species} of the corresponding quantum geon, and the remaining freedom
in the overall UIR of $G$ expresses the possibility of choosing an
arbitrary {\it statistics} (bose, fermi or para) for the geons of each
species.  For UIR's which represent the slides nontrivially, we do not
provide a complete classification, but we find some new types of effects
due to the slides, including quantum breaking of internal symmetry and of
particle indistinguishability.  In connection with the latter, a novel kind
of statistics arises which is determined by representations of proper
subgroups of the permutation group, rather than of the group as a whole.
Finally, we observe that for a generic 3-manifold there will be an infinity
of inequivalent UIR's and hence an infinity of ``consistent'' theories,
when topology change is neglected.


\sesquispace


\centerline {\bf  \S 1.  Introduction}
\Nobreak
\medskip
   
In any quantum theory of gravity based on a spacetime topology of the
product form\footnote{*}%
{In particular, canonically quantized gravity is included here, since it
seems necessarily to be based on a spatial manifold of fixed topology.}
$\R\times\M$, one is led naturally to the existence of distinct quantum
sectors labeled by the inequivalent unitary irreducible representations
(UIR's) of the group $G$ of large diffeomorphisms, or ``mapping class
group'', of the spatial 3-manifold $\M$ {\AmalfiBal} {\amalfi} {\rds}
{\isham} {\balrds}.  Indeed such inequivalent quantizations (or
``theta-sectors'') occur whenever the configuration space of a quantum
system has a non-trivial first homotopy group.  In the case of
(asymptotically flat) gravity on $\R\times\M$, the configuration space $Q$
can be taken to be the space of all 3-geometries on $\M$, and $\pi_1(Q)$ is
then isomorphic to the mapping class group (MCG) of $\M$.  In this paper we
analyze the UIR's of $G$ in a manner designed to bring out what the
different quantum sectors represent physically.

In \S 2 we review the structure of the mapping class group $G$ of a
three-manifold that is asymptotically ${\R}^3$.  Such a manifold can always
be expressed as a connected sum of ${\R}^3$ with prime manifolds.  We will
exclude handles from the primes, in which case $G$ is a semidirect product
of three subgroups (defined relative to a given presentation of the
connected sum): the ``slides'' of one prime through another, the
``permutations'' among identical primes, and the ``internal diffeos'' of
the individual primes.  In fact there is a triad  of normal subgroups,
  (slides) $\subseteq$ (slides and internal diffeos) $\subseteq$ G,
and associated with it, a semidirect product decomposition 
  $G$ = (slides) $\semidirect$ (internals) $\semidirect$ (permutations),
which is implicit in the literature.  Our demonstration of this semidirect
product form will use the concept of the ``development'' of a
diffeomorphism on one hand, and on the other hand, the fact that the MCG is
essentially identifiable with a subgroup of $\Aut(\pi_1(\M))$ whose
structure, in turn, can be deciphered with the help of the
Fuks-Rabinovitch presentation {\fuksrab} of the automorphism group of a
free product group.  Part of this demonstration is contained in the
Appendix.

In \S 3 we give a general analysis of the finite dimensional UIR's of
semidirect product groups.  Our exposition proceeds along the same lines as
that of {\mcka}, but differs from it at certain points because for
our purposes, one must consider projective representations of the normal
subgroup {\it without} relaxing the condition that the overall UIR one is
searching for be an ordinary (non-projective) representation of the full
group.  Full proofs for the structure theorems exposed in \S 3 may be found
in {\clifford} and {\thesis}.  In the subsequent sections of the paper, we
use the analysis of \S 3 to study the UIR's of the mapping class group of
$\M$.

The results of this study lend credence to the interpretation of quantized
primes as particles (topological geons).  Indeed, when the slide
diffeomorphisms are represented trivially, what remains is just a UIR of
the so-called ``particle-group'' (internals)$\semidirect$(permutations),
which describes the symmetry of a collection of particles with internal
structure.  For this group, {\it one is able to give explicitly the form of
the most general finite dimensional UIR}, and we do this in \S 6, using the
structure theorems of \S 3.  At the same time these structure theorems
provide a terminology which (remarkably enough) allows one to characterize
the corresponding quantum sectors entirely in the language of quantum
particles and their properties, providing in this sense a complete physical
interpretation for this class of UIR's.  Within this interpretation,
several general features emerge: the possibility of bosonic, fermionic and
para-statistics; the loss of the spin-statistics correlation; and the
quantum breaking of particle indistinguishability (i.e. the rendering
distinct of classically indistinguishable objects).

For general UIR's of $G$, we obtain (in \S\S 5 and 6) only partial results
because of our inability to categorize the UIR's of the slide subgroup.
However these results (illustrated with the example of ${\R}P^3$ geons)
already reveal several new effects associated with the slides, including
the breaking of the {\it internal} symmetry of individual geons, and the
rendering distinguishable of otherwise identical geons.  With three or more
geons the new effect emerges which is perhaps most striking: there are
sectors in which the particle statistics is dictated by representations,
not of the full permutation group, but by proper subgroups of it.  In
describing the effects in these terms, we are using the language
appropriate to the structure theorems of \S 3; however there seem to be
certain situations in which a slightly different language is more
appropriate, and we explore this possibility briefly in \S 6.

As we will see, quantum gravity without topology change manifests in
general an infinity of physically inequivalent quantum ``sectors'', an
embarrassment of riches which seems at odds with the conception of quantum
gravity as a fundamental theory.  In the concluding section we argue that
at a minimum, this indicates the necessity for incorporating topology
change within quantum gravity.  More realistically, it probably speaks to
the existence of an underlying discrete microstructure, for which continuum
quantum gravity can hope to provide only an effective low-energy
description.

\vskip 0.2 true in
\centerline {\bf \S 2. The Mapping Class Group: its Role and Structure}
\Nobreak
\vskip 0.1 true in

Let us first recall why the mapping class group (MCG) is relevant to
quantum gravity.  In an asymptotically flat product spacetime $\R\times\M$,
where the spatial $3$-manifold $\M$ is diffeomorphic outside of a compact
region to $\R^3$, the natural configuration space $Q$ of general
relativity is the space of all $3$-geometries on $\M$ which are
asymptotically flat.  Here a ``$3$-geometry'' is a diffeomorphism
equivalence class of Riemannian metrics on $\M$; however not all
diffeomorphisms should be considered ``pure gauge''  for this purpose.
Rather, since physical observables like momentum and angular momentum can
be expressed as surface integrals which generate diffeomorphisms at spatial
infinity, only those diffeomorphisms that vanish at infinity should be
considered gauge {\AmalfiBal} {\amalfi} {\balrds}.  From a less
formal point of view, we may say that a diffeomorphism which is non-trivial
at infinity actually effects a change in the relation of the isolated
system to its environment, and this is the physical reason why it would be
wrong to treat it as gauge.  Let $\Diff^{\infty}(\M)$ be the group of all
diffeomorphisms of $\M$ which vanish at $\infty$.  Then if ${\cal R}$ is
the space of all Riemannian metrics on $\M$ that are asymptotically flat,
we have $Q={\cal{R}}/\Diff^{\infty}$.  Since the action of $\Diff^{\infty}$
on ${\cal R}$ is free {\AmalfiBal} {\amalfi} {\balrds}, ${\cal R}$ is in
fact a principal fiber bundle with base space $Q$ and fiber isomorphic to
$\Diff^{\infty}$.

Fundamentally, the fact that $Q$, and not ${\cal R}$ as such, is the
natural configuration space of quantum gravity on $\R\times\M$, just
expresses the postulate of ``general covariance''.  However the formal
meaning of this fact appears differently in different formulations of
quantum gravity.  From the starting point of canonical quantization in the
Schr\"odinger representation, the so-called momentum constraints translate
into invariance under $\Diff^{\infty}_0$, the subgroup of $\Diff^{\infty}$
which is connected to the identity (also called the subgroup of ``small
diffeomorphisms'').  If implemented in the ``Dirac'' manner this invariance
asserts that the ``physical wave-functions'' are in effect functions only
over ${\cal R}/\Diff^{\infty}_0$ {\fsor}.  The further step of ``dividing
out'' by diffeomorphisms that are not in $\Diff^{\infty}$ must then be done
``by hand''.  From a path integral starting point, in contrast, both the
restricted and the full diffeomorphism invariances have clear geometrical
meanings.  Invariance under the spatial group $\Diff^\infty_0(\M)$ now
arises as a direct consequence of {\it spacetime} diffeomorphism
invariance; while the extension of this invariance group to
$\Diff^\infty(\M)$ occurs naturally as the result of summing over all
possible ways of ``attaching'' the spacetime manifold to the ``final
hypersurface'' $\M$, i.e., it arises as the consequence of a rudimentary
type of topology change.

The quotient group $\Diff^\infty/\Diff^\infty_0 =:\pi_0(\Diff^\infty)$ is
called the ``mapping class group'' (MCG) of $\M$; it is also often termed
the ``large diffeomorphism group''.  We shall denote it by $G$ and
sometimes call it simply ``the diffeo group''.  Given the fact that ${\cal
R}$ is convex and hence homotopically trivial, the homotopy exact sequence
for a principal fiber bundle leads to the isomorphism,
$\pi_1(Q)\simeq\pi_0(\Diff^{\infty})=G$.

Now, it is well known that the existence of a non-trivial fundamental group
of the configuration space leads to inequivalent quantizations determined
by the UIR's of that group.  In the canonical framework, 
 each inequivalent ``sector'' of Hilbert
space is realized in terms of an $n$ dimensional vector-bundle over $Q$,
where $n$ is the dimension of the corresponding UIR of $\pi_1(Q)$ and the
state vectors or ``wave functionals'' are the cross sections of the vector
bundle. 

The role of the MCG emerges even more directly from a ``path-integral'' or
``sum-over-histories'' starting point.  In that framework the fundamental
dynamical input is a rule attaching a quantum amplitude to each pair of
truncated histories which ``come together'' at some ``time'' {\unimod}
{\drexel}.  Let us call such a pair a ``Schwinger history'' for short, and
its underlying manifold a ``Schwinger manifold''.  In the case of quantum
gravity, each separate truncated history is a Lorentzian manifold with
final boundary (and possibly initial boundary depending on the physical
context), and the ``coming together'' means the identification or ``sewing
together'' of the final boundaries.  Now, as alluded to above, different
ways of sewing are possible, related to each other by large diffeomorphisms
of the final boundary.  (This final boundary corresponds to the spacelike
slice $\M$ of the canonical formulation).  In general such a
re-identification may or may not lead to a diffeomorphic Schwinger
manifold, but it never will if we restrict ourselves to product spacetimes
of the form $\R\times\M$, i.e. if we exclude topology change from the
truncated histories (and if we limit ourselves to diffeomorphisms vanishing
on any initial boundaries which may be present).  In this case, the MCG of
$\M$ acts freely and transitively (albeit non-canonically) on the set of
Schwinger manifolds.  Now, without disturbing the classical limit of the
theory or the local physics, we can multiply the amplitude of each
Schwinger history by a complex number depending only on the topology of the
underlying manifold.  Somewhat analogously to {\cecile}, one can then argue
that consistency requires that these complex weights transform under some
unitary representation of $G$, and that weights belonging to disjoint
representations of $G$ ``do not mix''.  The pure cases are then the UIR's
of $G$, and we arrive again at the conclusion that each distinct UIR of the
MCG yields an inequivalent version or ``sector'' of quantum gravity without
topology change.  Thus a study of the UIR's of the diffeo group is required
to understand the possible inequivalent sectors of quantum gravity
{\AmalfiBal} {\amalfi} {\balrds}.

[ In the context of canonical quantization, two related questions might
arise at this point, one concerning the role of the spacetime
diffeomorphisms, as opposed to the purely spatial ones, and the second
being the question why we should be dealing only with UIR's of the MCG,
rather than general UIR's of the full invariance group $\Diff^\infty$.
From the point of view of the wave-function, the spacetime diffeos do not
appear as such, but, as is well known, their influence can be felt in a
further set of restrictions on the wave-function, the ``Hamiltonian
constraints''.  (The precise content of these constraints depends on
whether or not one adopts the so-called ``unimodular'' modification of
general relativity {\unimod}.  If one does not adopt it, then there is, so
to speak, one constraint for each point of the 3-manifold $\M$.  If one
does adopt it then one of these constraints disappears and a Schr\"odinger
equation takes its place.)  Unlike with the passage from
$\Diff^\infty_0(\M)$ to $\Diff^\infty(\M)$, it does not seem possible to
encode this infinity (or ``infinity minus one'' in the unimodular case) of
constraints into a further contraction of the effective configuration space
$Q$.  On the other hand, the presence of this ``extended gauge invariance''
should probably be understood as the reason why it is only the quotient
group $\pi_0(\Diff^\infty)$ (and not $\Diff^\infty$ itself), whose
representations we need concern ourselves with. ]

For a generic 3-manifold the mapping class group is a cumbersome group to
deal with since it is both infinite and discrete, and such groups tend to
possess representations of types II and III, unless they are ``almost
abelian''.  (For example, the regular representation of an infinite
discrete group $G$ is type II if the quotient of $G$ by the subgroup of all
elements belonging to finite conjugacy classes is infinite {\mcka}.)  But
type II and III representations have the unsettling property of a
non-unique decomposition into irreducibles, which would hamper any attempt
to analyze the physical behavior of geons described by such representations
solely in terms of UIR's of $G$.  There is also the worry whether the
infinite multiplicity of states associated with an infinite dimensional
representation of $G$ can be physically acceptable.  To avoid such
problems, we will restrict ourselves in this paper to finite dimensional
representations of $G$, for which the decomposition into UIR's is always
unique.  In addition the restriction to finite dimensionality will greatly
simply the general analysis of \S 3 below.

The simplest example of the influence of the MCG is the existence of
nontrivial $2\pi$ rotations of the three manifold $\M$, implying the
possibility of half-integer angular momentum in pure gravity {\fsor}
{\hriks}.  Since the 4$\pi$ rotation {\it is} trivial, one has only the two
UIR's of the group ${\Z}_2$, and hence the choice between spinorial and
tensorial sectors.  Given the whole diffeo group of the three manifold
then, it is natural to inquire whether other kinds of interesting behavior
may be found, but before addressing this question, we must examine the MCG
itself more closely.
 

\noindent {\bf The structure of the mapping class group}
\par\nobreak

Let us consider first the case of oriented three-manifolds.  Then
{\hempel}, any manifold $\M$ that is diffeomorphic to $\R^3$ outside of a
compact region can be decomposed uniquely as a connected sum of $\R^3$ with
closed prime manifolds, which in this sense are the fundamental
constituents of $\M$.  We write this decomposition as
$\M=\R^3\#(\#_{i}{\cal P}_i)$, where the connected sum operation $\#$ is
defined as follows.  If $X$ and $Y$ are oriented three-manifolds, remove a
3-ball (or ``disk'') $\D^3$ from each of them, to get $X\backslash
\D^3$ and $Y\backslash \D^3$, thus creating an $S^2$ boundary for each;
then identify along these boundaries in an orientation-consistent manner to
get the oriented manifold $X\# Y$.  By a prime manifold ${\cal{P}}_i$ we
mean, then, a closed three-manifold that cannot be further decomposed into
a connected sum of other three manifolds.  (The only possible further
decomposition is trivial, and is a connected sum of the prime itself with a
three sphere $S^3$, ${\cal P}_i={\cal P}_i \# S^3$.)  For some
examples of prime manifolds see {\fwit}.  We will refer to the quantized
primes as {\it topological geons}, or just ``geons'' for short.

For unoriented manifolds (including unorientable ones), the situation is
similar but less simple to express, because of two kinds of ambiguity which
can be present.  First, the connected sum itself can be ambiguous, because
there are two distinct ways to identify the bounding 2-spheres of the
excised $\D^3$'s, related to each other by a parity flip.  When both $X$
and $Y$ are {\it chiral}, the two identifications will yield inequivalent
(non-diffeomorphic) results for $X{\#}Y$.  (A chiral manifold is an
orientable manifold that admits no orientation-reversing diffeomorphism.)
Thus, in forming an unoriented connected sum whose summands include $n$
chiral primes and no non-orientable primes, one has $2^{n-1}$ possible
outcomes.\footnote{*}
{In the asymptotically flat setting, we may regard $\R^3$ as a chiral
prime.  Then a sum of it with $n$ chiral (closed) primes can performed in
$2^n$ different ways.}
In the geon language, one has a choice of whether each prime will occur as
a certain geon or as the $CP$--conjugate geon (antiparticle).  Notice that
this ambiguity disappears if even one non-orientable prime is present.  It
would thus disappear entirely if we limited ourselves to manifolds which
were either oriented or not orientable at all.  The second kind of
ambiguity is that the decomposition into primes is in general non-unique
when handles are present.  Specifically, if $\M$ is non-orientable then we
have the isomorphism
$$
  S^2{\times}S^1 \ \# \ \M  \isom  S^2{\tilde\times}S^1 \ \# \ \M,
$$
where $S^2 \times S^1$ is the ``orientable handle'', and
$S^2{\tilde\times}S^1$, its $\Z_2$--twisted analog, is the ``non-orientable
handle''.  Thus, the two types of handle are interchangeable summands in
certain situations.  However, we will be excluding handles from
consideration altogether, and so the minor non-uniqueness they entail will
be of no consequence for this paper.

We now exclude the handles by specializing to the case where $\M$ is a
connected sum of irreducible primes, these being the closed three manifolds
for which any embedded two-sphere $S^2$ can be shrunk to a point.  The only
orientable prime three-manifold which is not irreducible is the $S^2\times
S^1$ handle.  (Sometimes called ``wormhole'', this is probably the best
known --- if least generic --- of all primes).  The diffeomorphism group of
a manifold containing such handles exhibits some peculiarly nonlocal
behavior that interferes with a particle-like interpretation of handle
primes {\mcmil} {\nico} {\cox}.  Possibly, one could attain a consistent
particle interpretation by treating the ends of a handle as separate
entities in certain situations (i.e. in certain regions of state space),
but we will not explore that possibility herein.  Instead, we will simply
limit ourselves to manifolds $\M$ not containing summands $S^2{\times}S^1$
or $S^2{\tilde\times}S^1$.  Notice that with the exclusion of handles, the
prime decomposition of $\M$ becomes unique, even in the unoriented
case.

Having excluded handles, we now also limit ourselves to orientable
manifolds $\M$.  Unlike with the handles, we know of no impediment to
extending our analysis to the non-orientable case.  On the contrary, there
are interesting new features (like interconversion between particles and
anti-particles) which seem to fit well into our scheme.  Rather we exclude
nonorientable primes only because, at several points, we use theorems whose
nonorientable extensions we have not been able to find in the literature.
However, we believe these extensions exist (mutatis mutandis), and we will
at various points comment on the nonorientable case, always with the
proviso that what we say depends on the continued validity of certain
theorems known to obtain in the orientable case.

Let us fix a concrete representation of $\M$ as a connected sum of primes
(a ``presentation'').  It is well documented in the mathematics literature
{\mcmil} {\hrksmc} {\mcc}, that with respect to such a presentation, the
MCG of $\M$ is generated by the following elements, a description of which
can be found in {\amalfi}.

\item {(a)} internal diffeos of the individual primes

\item {(b)} exchanges of identical primes

\item {(c)} slides of primes through other primes along generators of
              $\pi_1(\M)$.

Let us examine the structure of the MCG more closely, in the case in which
$\M$ is a connected sum of $N$ identical primes ${\cal{P}}_i$.  (The
general case is more complicated only in notation.)  For our subsequent
analysis the important circumstance will be that $G$ is the semidirect
product of the three subgroups generated by the three types of
diffeomorphisms just enumerated.\footnote{*}
{It is this semidirect product decomposition that fails when handles are
present.}
We will call these respectively the subgroup $G_{int}$ of internal
diffeomorphisms, the permutation subgroup $S_N$, and the slide subgroup
$\cS$.  In the next few paragraphs we discuss the structure of these three
subgroups, and we describe the manner in which $G$ is their semidirect
product.  Some of the relevant facts we will state without proof, giving
demonstrations only insofar as they seem helpful for arriving at a physical
interpretation of the various elements of $G$ and its UIR's.  In the
Appendix, we describe how to make the proofs complete (assuming the
Poincar\'e conjecture to be true.)

The subgroup of {\it internal diffeomorphisms} (the subgroup generated by
the diffeomorphisms of type (a)) is the direct product
\eqn\gint
{
  G_{int} = G_1 \times G_2 \times \cdots \times G_N ,
}
where $G_i$ is the MCG of $\R\#\cP_i$, and the $G_i$ are mutually
isomorphic, because all the primes are identical.  The product is direct
since the support of each $G_i$ is by definition restricted to the region
inside the $i^{th}$ separating sphere (which as its name indicates,
separates the $i^{th}$ prime from the rest of $\M$), implying that the
internal diffeos corresponding to different primes do not mix (see the
Appendix).

The {\it permutation subgroup} (the subgroup of $G$ generated by
diffeomorphisms of type (b)) consists of the $N!$ diffeomorphisms that
permute the $N$ identical primes ${\cal P}_i$ amongst themselves.  (When
$\cP$ is nonorientable, we assume for purposes of defining the exchanges
that the presentation of $\M$ has been chosen so that all the $\cP_i$ are
diffeomorphic via translation through the ambient $\R^3$.)  Thus (see the
Appendix) its structure is just that of the permutation group on $N$
elements, and we have given it the same name, $S_N$.

The {\it slide subgroup} $\cS$ (the subgroup of $G$ generated by
diffeomorphisms of type (c)) is the free product
$$
       {*}_i \ {*}_j \  {\rm S}_{ij}  \qquad (i,j=1\cdots N,\ i\not=j)
$$
modulo certain geometrically evident commutation relations, where
${\rm{S}}_{ij}\isom\pi_1(\cP_j)$ comprises the slides of the $i^{th}$ prime
through the $j^{th}$ prime (that is, an element of ${\rm{S}}_{ij}$ slides
$\cP_i$ along a loop belonging to $\pi_1(\cP_j)$).  When only two primes
are present $\cS$ is literally the free product of ${\rm S}_{12}$ with
${\rm{S}}_{21}$.  However when three or more primes are present certain
commutation relations among the generators of distinct subgroups must be
imposed, as described in {\fuksrab} {\mcc}.  For example for $N=4$,
${\rm{S}}_{12}$ obviously commutes with ${\rm S}_{34}$.  A complete set of
these relations is as follows:
$$
	{\rm s}_{ij} \,\natural\, {\rm s}_{kl} ,  \qquad
	{\rm s}_{ij} \,\natural\, {\rm s}_{kj} , \qquad
	{\rm s}_{ik}{\rm s}_{jk} \,\natural\, {\rm s}_{ij} ,
$$
where $\{i,j,k,l\}$ label distinct primes, ${\rm s}_{ij}$ is any generator
of ${\rm S}_{ij}$, and the symbol $\natural$ denotes commutability:
$$
      A \,\natural\, B \ideq AB=BA .
$$  

Generators for the groups ${\rm{S}}_{ij}$ can be chosen as follows.  Let
$\gamma_i^{\phi}$ be generators of $\pi_1({\cal P}_i)$ where $\phi$ runs
from $1$ to $n$.  Clearly there will be $n\times N$ such $\gamma_i^{\phi}$
in all.  Now for each such generator passing through the $i^{th}$ prime,
there is a slide of any one of the other primes along it.  Let $s_i^{\phi,
j}$ denote the slide of the $i^{th}$ prime along the $\phi^{th}$ generator
of $\pi_1$ of the $j^{th}$ prime.  There will be $nN(N-1)$ such generators
in all, since a prime cannot slide through itself.  (In the non-orientable
case, some of these generators can fail to exist, when more than one type
of prime is present.  Thus, if one attempts to slide a chiral prime $\cP_i$
along an orientation reversing loop $\gamma\subseteq\M$, one fails because
$\cP_i$ returns as the mirror image of itself, which (in the language used
below) precludes the would-be ``development loop'' from closing.  Hence,
the group defined by the generators and relations just described contains
spurious elements, and the actual $\slides$ is a subgroup of it, namely the
subgroup of elements which contain an even number of orientation reversing
generators $s_i^j$ for each chiral prime $\cP_i$.) 

A second subtlety concerning the slides involves spinorial primes (those
whose $2\pi$-rotation is nontrivial).  Above we wrote that the generators
(a)--(c) were defined {\it relative to} a presentation of $\M$ as a
connected sum.  We note here that such a presentation involves more than
just a choice of separating $S^2$'s and of identifications among the
identical primes.  It requires as well a distinguished notion of ``parallel
transport'' of the primes to enable one to disentangle the ``pure slides''
from ``slides mixed with $2\pi$-rotations of the sliding primes''.  For a
fuller discussion of this point see {\balrds}.  (In the non-orientable case
one can transport chiral primes with respect to the orientable double cover
of $\M$.  We believe a similar technique should work also for the transport
of nonorientable primes and non-chiral orientable primes.)

Given the slides, the internal diffeos and the permutations, one can build
up the entire mapping class group $G$ as the semidirect product
\eqn\comp
{
    G =  \cS \semidirect   G_{int}  \semidirect S_N
}
(By writing $G=A\semidirect B\semidirect C$ we mean the following.
Every $g\in{}G$ is uniquely of the form $g=abc$ with
$a\in{}A$, $b\in{}B$, $c\in{}C$; and for all such $a$, $b$, $c$, we have 
$bab^{-1}\in{A}$, $cac^{-1}\in{A}$ and $cbc^{-1}\in{B}$.  This is
equivalent to  $G=A\semidirect(B\semidirect C)$ and implies 
$G=(A\semidirect B)\semidirect C$.)  The proof of this semidirect product
structure is described in the Appendix, relying heavily on the device of
replacing the study of $G$ as such by the study of its action on
$\pi_1(\M)$.  Here we limit ourselves to a partial treatment intended to be
useful in gaining an intuitive understanding of the structure of $G$.

First, let us show that the slide subgroup ${\cal S}$ is normal in $G$, as
required by the decomposition {\comp}.
Consider the slide $s$ of a prime ${\cal P}_i$ along a generator of
$\pi_1(\M)$ that passes through the prime ${\cal P}_j$, say.  This slide
will commute with all permutations of the primes except those involving
${\cal P}_i$ and ${\cal P}_j$, and with all internal diffeos but those of
${\cal P}_j$.  When conjugated by permutations involving ${\cal P}_i$ or
${\cal P}_j$, $s$ becomes another slide, now between the permuted primes.
Thus, if we can show that conjugation of $s$ by an internal diffeo of
${\cal P}_j$ also yields another slide, then we can state that the slides
form a normal subgroup in the diffeo group.  We proceed to do this using
the idea of the development of a diffeomorphism.

A diffeo is said to be {\it developed} in the following sense {\amalfi}.
If you start off with the manifold $\M$ and ``implement the diffeomorphism
in a continuous manner'', then you obtain a sequence of manifolds that are
diffeomorphic to $\M$ but {\it different}, except when the diffeo is
completed, in which case one is back to the same manifold as before.  That
is, the development is a {\it loop of manifolds} that are diffeomorphic to
each other, with the base point being $\M$, and such that in going around
the loop, one has in effect performed the above diffeo.  For example, the
development of a diffeomorphism exchanging two $T^3$ primes is illustrated
in figure 1.

\vbox
 {\bigskip
 \centerline {\epsfbox{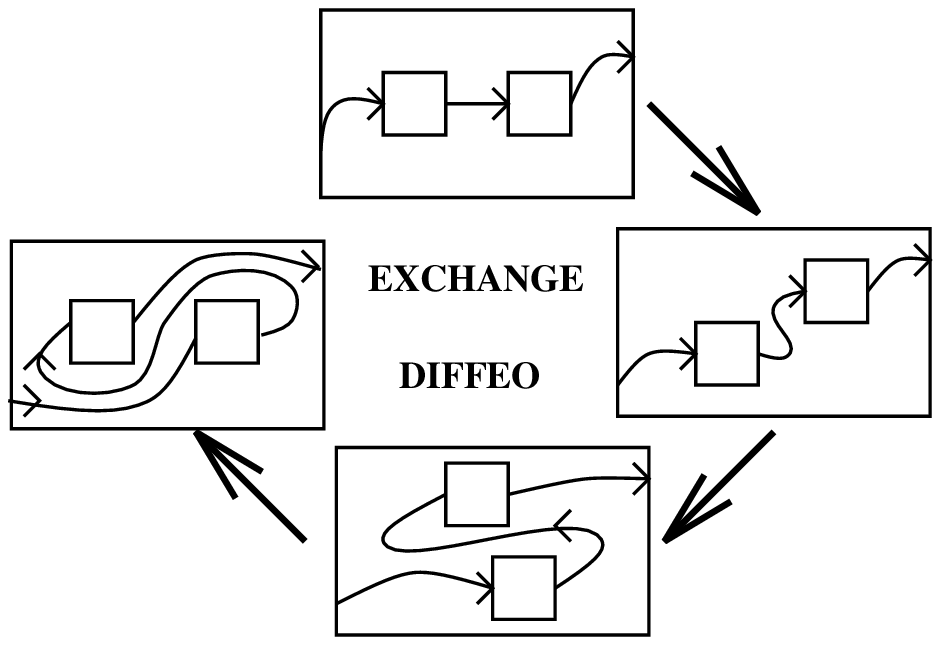}}
 \bigskip
 \noindent  
 \singlespace
 \noindent{\bf Figure 1.}
   {\bf  Development of the exchange diffeomorphism for two $T^3$ primes.}
   The line with arrows shows the effect of the
   diffeomorphism on a curve whose end points are fixed at infinity and
   which loops through the two primes non-trivially. }
\bigskip

The diffeo we are interested in is $gsg^{-1}$ where $g$ is the internal
diffeo and $s$ is the slide diffeo.  So we develop $g^{-1}$ first, then $s$
and then $g$.  This gives us a single loop of manifolds, as illustrated in
figure 2.  The aim now is to show that this loop is in fact deformable to
one that develops just a slide.  To do so, we can proceed as follows.
Instead of completing all of $g^{-1}$ we leave a little bit of this
diffeomorphism undone (hence the manifold thus obtained is {\it different}
from the one we started out with).  The slide is now developed by taking it
along the ``same'' generator as before.  The remaining bit of $g^{-1}$ is
then completed and finally $g$ is developed exactly as before.  The result
is a slightly different loop of manifolds from originally.  Now we keep
decreasing the amount of $g^{-1}$ that is developed before the slide is
executed so that we obtain a whole sequence of loops parameterized
continuously by some $\epsilon \in [0,1]$ which measures the ``amount'' of
$g^{-1}$, ``$\epsilon g^{-1}$'', that has been left undone before the slide
is executed.  The final loop in this sequence is one for which none of
$g^{-1}$ is done before the slide, and the slide is along a new generator
obtained by continuity from the original one.  Of course, developing
$g^{-1}$ and $g$ consecutively gives the identity, and thus this last loop
involves only a slide along a generator that is different from the one for
the original slide.

\bigskip
\vbox{
 \centerline {\epsfbox{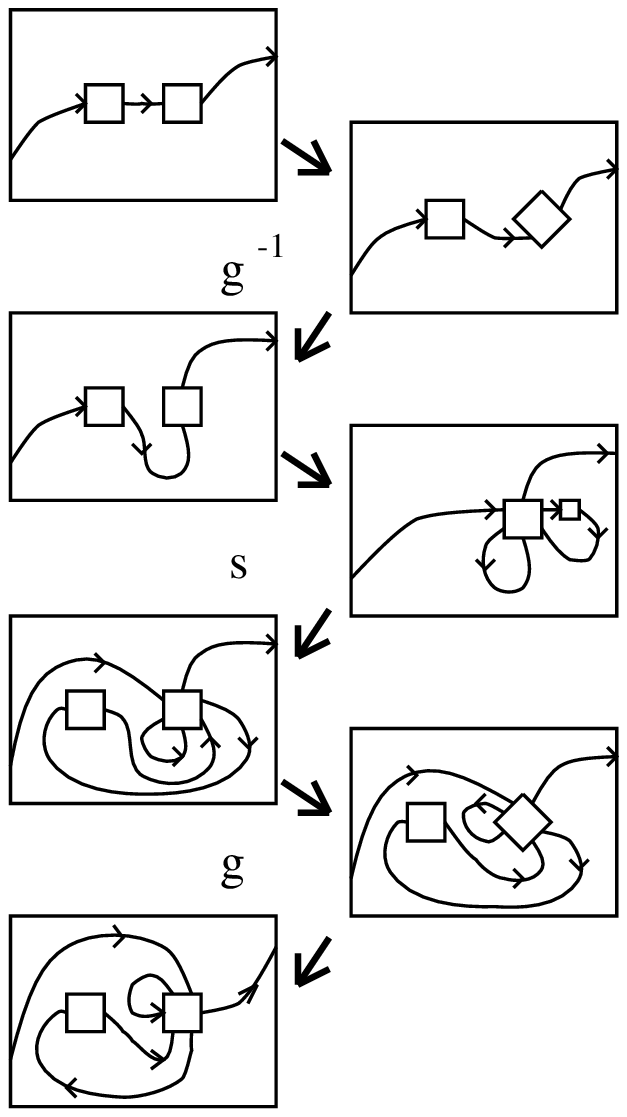}}
 \bigskip
 \singlespace
 \noindent{\bf Figure 2.} 
 {\bf Development of the diffeomorphism $gsg^{-1}$ for two $T^3$ primes.} 
  Here $g$ is an ``internal rotation'' by $\pi /2$, and $s$ is a slide of
  the first prime through the second.  Panels 1--3 develop $g^{-1}$, 3--5
  develop $s$, and 5--7 develop $g$.}
\bigskip

A generic loop in our sequence of developments can be described in the
following way.  Perform $\epsilon g^{-1}$ first, then slide ${\cal P}_i$
along the same generator of ${\cal P}_j$ that it goes through when
$\epsilon=1$.  Finish the development by doing $(1-\epsilon)g^{-1}$ and
then $g$.  Since these loops are all deformable into each other, the
diffeomorphisms they develop are isotopic to each other: they yield the
same element of $G=\pi_0({\Diff^{\infty}})$.  Thus we have shown that
conjugation of a slide diffeo by an internal diffeo of the prime slid
through is a slide along another generator of the same prime.  Hence the
slide subgroup is normal in the full mapping class group $G$.

In the same way, it is clear that for any $g_{int}\in{}G_{int}$ and any
$p\in{}S_N$, $pg_{int}p^{-1}$ is again an internal diffeo.  For example,
imagine that $p$ interchanges the $i^{th}$ prime with the $j^{th}$ one and
that $g$ is an internal diffeomorphism of the $i^{th}$ prime.  Then
conjugation of $g$ with $p$ gives the corresponding internal diffeo of the
$j^{th}$ prime.  Thus we have demonstrated everything needed for {\comp}
but the uniqueness of the decomposition.  This is done in the Appendix.

Finally, consider the quotient group $G/{\cal S}$, which by {\comp} may be
identified with the subgroup $\tG \subseteq G$ generated by the elements of
$S_N$ and $\capG$ (i.e., $\tG=G_{int}S_N$).  This group has been dubbed
{\it the particle group}, since its elements have natural interpretations
as symmetry operations on particles possessing internal structure.  By
{\comp}, it also is a semidirect product, and this fact will be the key,
both to our analysis of its representations and to their physical
interpretation.

The important structural result of this section is that $G$ can be
expressed as a semidirect product of three subgroups with clear physical
meanings.  Since we are interested in understanding the finite dimensional
UIR's of $G$, it is therefore important to understand in general the
structure of the UIR's of semidirect product groups.  To that end, we
provide the analysis of the next section.

\vskip 0.2 true in


\centerline
  {\bf \S 3. The Structure of Representations of Semidirect Product Groups}
\Nobreak
\vskip 0.1 true in

Consider a group $G$ with normal subgroup $N$ and quotient group $K=G/N$.
In the language of exact sequences we express this situation as
\eqn\eseq
{
  1 \rightarrow N \rightarrow G \rightarrow K \rightarrow 1
}
We will assume in addition that this sequence {\it splits}, meaning that
its projection-homomorphism, $\pi : G\rightarrow K$ possesses a ``section''
or right inverse, i.e., that there exists a homomorphism $j:K\rightarrow G$
such that $\pi j=1_K$.  In less poetic language what we are saying is that
$G$ can be realized as a semi-direct product, $G=N \semidirect K$, which we
do by using $j$ to construe $K$ as a subgroup of $G$ complementary to $N$.
The precise realization of $K$ as a subgroup of $G$ depends on the
splitting homomorphism $j$ , but we will choose one such $j$ and never
change it.  We then have the situation:
\eqn\subset
{
  N\subseteq G,  \   K\subseteq G,  \   N\cap K=\{e\}, \  G=NK,
}
where $e$ is the identity element of $G$. (The ambiguity in $j$ is not an
irrelevant technicality for our application to geons: it corresponds to the
ambiguity in the presentation of $\M$ as a connected sum, and is linked
intimately to the subtle distinction between concrete primes and the
corresponding physical particles (geons).  See {\amalfi} and {\rds}.)

We aim to understand the general finite dimensional unitary
irreducible representation (UIR) of G, by analyzing it in terms of
representations of $N$ and representations of (the appropriate
subgroups of) $K$.  In fact one can begin with an arbitrary UIR $R:
G\rightarrow Aut({\cal H})$ and analyze it into certain structures
involving $N$ and $K$.  Conversely one can show that, beginning with
these structures one can build up a representation of $G$ in a unique
manner.  These two inverse processes are described fully in {\thesis};
here we merely summarize their main features.  References {\clifford}
and {\mcka} may also be consulted for much of this material, including
an explanation of any terminology not defined here.

  In what follows we use the symbol $\simeq$ to denote isomorphism of
groups, vector spaces or group representations as the case may be.
Similarly $\otimes$ will stand for several types of tensor product: of
vectors, operators or representations.  In the case of representations we
make the following convention.  If $R_1$ and $R_2$ are representations of
{\it different} groups, $G_1$ and $G_2$, then $R_1\otimes R_2$ will denote
the tensor-product representation of the direct product group
$G_1\times{}G_2$.  However, when $G_1=G_2=G$, then $R_1\otimes R_2$ will
denote not a representation of $G\times G$, but the tensor product
representation of $G$ itself (the latter or ``inner product''
representation of $G$ being just the restriction of the former or ``outer
product'' representation of $G\times{}G$ to the diagonal subgroup within
$G\times G$).  In each case the context should make clear which product is
intended.

\noindent {\underbar{The ``Little Group'' and Associated Structures}}

Before proceeding to the analysis of $R$, we must expose some structures
and relations which are available independently of reference to any
particular UIR of $G$.  To begin with, let $\hN$ be the set of (finite
dimensional) UIR's of $N$, up to equivalence; i.e., an element of $\hN$ is
an equivalence class of concrete representations $\Gamma$ of $N$, where
${\Gamma}_1\simeq {\Gamma}_2 \iff \exists$ some unitary $S$, such that
$\forall$ $n\in N$,
\eqn\simequal
{
   S \, {\Gamma}_1(n)={\Gamma}_2(n)S
}

Because $N$ is normal in $G$, $K$ acts naturally on $\hN$, the action of
$k\in K$ on the UIR ${\Gamma}$ of $N$ being to produce the UIR ${\Gamma}'$
defined by
\eqn\threea
{
  {\Gamma}'(n)={\Gamma}(knk^{-1}), \forall n \in N.
}
We will express this relationship by writing ${\Gamma}'={\Gamma}k$, placing
$k$ on the right since our definition of ${\Gamma}'$ yields a {\it{right}}
action in the sense that $({\Gamma}k_1)k_2 ={\Gamma}(k_1k_2)$.  Since
${\Gamma}_1\simeq{\Gamma}_2\Rightarrow{\Gamma}_1k\simeq{\Gamma}_2 k$, we
acquire an action $[\Gamma]\mapsto[\Gamma]k$ of $K$ on $\hN$ (where
$[{\Gamma}]$ means the equivalence class in $\hN$ of ${\Gamma}$).

Now, let $\cO \subseteq \hN$ be an {\it orbit} of the $K$-action just
defined.  If we choose a fiducial element $[{\Gamma}_0]\in \cO$, then we
can define the associated ``little group'' $K_0$ to be the ``stability
subgroup'' of $K$ with respect to $[{\Gamma}_0]$:
\eqn\stability 
{
K_0=\{ k\in K | [{\Gamma}_0]k=[{\Gamma}_0]\}
}
In the following we will assume that such a fiducial element has been
chosen, once and for all, in every orbit $\cO$.

If, further, we choose a fiducial {\it representative},
${\Gamma}_0\in[{\Gamma}_0]$,  then we acquire a {\it projective unitary
representation} of $K_0$ as follows.  By definition,
${\Gamma}_0{k}\simeq{\Gamma}_0$ for $k\in K_0$. But this means
\eqn\threeb
{ 
  {\Gamma}_0(knk^{-1}) = S(k) {\Gamma}_0(n) S(k)^{-1}
}
for some unitary operator $S(k)$ acting in the carrier space of
${\Gamma}_0$.  Since $n$ in {\threeb} is arbitrary, $S(k)$ is thereby
determined uniquely up to phase.  Compatibility with group multiplication
in $K$ then implies that
\eqn\threec
{
  {S(k_1)S(k_2)=\sigma (k_1, k_2) S(k_1 k_2)},
}
where $\sigma : K_0 \times K_0 \rightarrow U(1)$ is a {\it projective
multiplier} for $K_0$.  We remind the reader that a fixed, concrete
${\Gamma}_0$ has been chosen for every orbit, and we hereby choose also a
fixed set of phases for the $S(k)$.  (Of course we will choose these phases
so that $\sigma\equiv 1$ when that is possible).

\noindent 
{\bf To summarize}: Given the split exact sequence
$1 \rightarrow N \rightarrow G \rightarrow K \rightarrow 1$ 
(together with choices of certain fiducial elements where necessary) we
acquire automatically: 

\item {$\bullet$} an {\it action} of $K$ on the UIR's of $N$;  and for each
orbit $\cO$ of this action in $\hN$:  

\item{$\bullet$} the corresponding {\it stability group} 
      (or ``little group'') $K_0 \subseteq K$; 

\item {$\bullet$} a {\it projective multiplier} $\sigma$ for $K_0$; and

\item {$\bullet$}  a {\it projective representation} $S$ of $K_0$ with
multiplier $\sigma$ (with $S$ acting in the carrier space ${\cH}_0$ of the
fiducial representation ${\Gamma}_0$ chosen for the orbit $\O$).

To acquire these structures we have had to choose certain fiducial
elements, and one may be curious how an altered choice would affect the
above acquisition list.  The choices made comprise: a choice of splitting
map $j\,;$ and for each orbit $\cO$~: a choice of UIR ${\Gamma}_0$ with
$[{\Gamma}_0]\in \cO$  plus a choice of phase for each operator $S(k)$.
Changing the splitting map $j$ alters the subgroup of $G$ which represents
$K=G/N$, but it does not affect the action of $G/N$ on $\hN$.  It also need
not affect the projective multiplier $\sigma$, although it does necessarily
modify (in a simple manner) the operators $S(k)$.  Changing the fiducial
element $[{\Gamma}_0]\in \cO$ changes $K_0$ by an inner automorphism of
$K$, and both $S$ and $\sigma$ can be given the new values induced by this
inner automorphism; in that sense they remain unaffected.  Changing which
${\Gamma}_0$ represents the fiducial element of $\cO$ need have no effect
other than a similarity transformation on the operators $S(k)$.  Finally,
changing the phases of the $S(k)$ modifies only $\sigma$, taking it by
definition to another member of the same projective equivalence class
$[\sigma]$ (i.e. another cocycle representing the same element of the group
cohomology of $K_0$).

\noindent {{\underbar {Analysis of a UIR of G}}}

Let $R:G\rightarrow Aut(\cH)$ be a finite dimensional unitary irreducible 
representation (UIR) of $G=N
\semidirect K$ acting in the Hilbert space $\cH$, and let $\tR=R | N$
be its restriction to the normal subgroup $N$ (i.e. $\tR: N
\rightarrow Aut(\cH)$ via $\tR(n)=R(n)$).  The analysis of $R$ then
runs as follows.

First, $\tR$ must (up to a similarity transformation) take the form 
\eqn\threed
{
  \tR := R|N=(\bigoplus_{[{\Gamma}] \in\cO}{\Gamma}) \otimes {\bf 1}_\beta,
}
where $\cO \subseteq \hN$ is some definite $K$-orbit (determined by $R$)
and ${\bf 1}_\beta$ denotes the trivial representation of $N$ in
$\cC^{\beta}$.  Here, the direct sum notation denotes a sum of UIR's taken
over the orbit $\cO$.  That is, we select from each equivalence class
${\alpha}_i\in \cO$ ($i=0, \cdots, \hbox{$|\O|-1$}$) a representative
representation ${\Gamma}_i\in{\alpha}_i$, and we form the direct sum
representation ${\bigoplus}_{i=0}^{|\cO|-1}{\Gamma}_i$.  For $i=0$ we will
make the convention that ${\alpha}_0$ is the fiducial equivalence class in
$\cO$, and we of course use the previously selected fiducial ${\Gamma}_0\in
{\alpha}_0$ as ${\Gamma}_i$ when $i=0$.

Now let $K_0$ be the little group of $\cO$, as always, and let
$S:K_0\rightarrow Aut(\cH_0)$ be the corresponding projective
representation of $K_0$ with multiplier $\sigma$, as described above.  From
{\threed}, one component of $\tR$ is the representation ${\Gamma}_0 \otimes
{\bf 1}_{\beta}$ acting in the subspace $\cH_0\otimes \cC^{\beta}$.  Then
the restriction of the full representation $R$ to $K_0$ induces {\it in
this subspace} a representation of the little group $K_0$ which has the
form,
\eqn\threee
{
  S\otimes T,
}
where $T$ is some projective unitary irreducible representation (PUIR) of
$K_0$ with multiplier $\tau=1/\sigma$.  The representation $T$ and the
orbit $\cO$ together contain enough information to identify uniquely the
representation $R$ from which they came.

Note here that $T:K_0 \rightarrow Aut(\cC^{\beta})$ is not merely a
projective equivalence class, but is a concrete collection of operators
$T(k)$.  Unfortunately, it too is not uniquely determined, but depends on
the presentation of $\cH_0\otimes\cC^{\beta}$ as a tensor product.
However a change of this presentation will only induce a similarity
transformation, $T(k)\mapsto UT(k)U^{-1}$.  Notice also, that a change
to the phases chosen for the $S(k)$ would not alter the projective
equivalence classes of $T$ and $\tau$.  In fact a phase change which just
multiplied $S(k)$ by a unitary {\it character} $\chi(k)$ of $K_0$ would not
alter $\tau$ at all (since it would not alter $\sigma$).

\noindent 
{\bf To summarize}: We have ``analyzed'' $R$ into a pair of structures: an
orbit $\cO\subseteq\hN$, and a PUIR \ $T$ of the corresponding little group
$K_0$ with multiplier $1/\sigma$ ($\sigma$ being the projective multiplier
associated to $\cO$ as described as above).  $T$ is determined by $R$ up to
a similarity transformation, $T \rightarrow UTU^{-1}$.

\noindent {\underbar {Synthesis of a UIR of G}}

Now let us retrace our steps.  Given any finite $K$-orbit $\cO\subseteq\hN$
and any finite dimensional PUIR $T : K_0\rightarrow Aut(\cH_T)$ of the
corresponding little group with projective multiplier $\tau=1/\sigma$, we
can construct a finite dimensional UIR $R$ of $G$.  In this process we
begin by representing $N$ and $K_0$ jointly in a certain ``core'' Hilbert
space, and then we build up the rest of $\cH$ by acting on the core with
elements of $K/K_0$.  

Specifically, let $\cH_0$ be the carrier space of ${\Gamma}_0$ and let
$\cH_T$ be that of $T$ (so $\cH_T\simeq\cC^{\beta}$ where $\beta=dim T$).
Then our ``core'', is the tensor product space,
\eqn\hilbert
{
     \cH_0\otimes \cH_T
}
and our representation of $NK_0$ is that for which $n \in N$ acts as
${\Gamma}_0(n)\otimes {\bf 1}$ and $k \in K_0$ acts as $S(k)\otimes T(k)$.

So far we have a UIR of $NK_0\subseteq G$ in $\cH_0 \otimes \cH_T$.  The
full carrier space (or ``substrate'') ${\cH}$ of $R$ is then built up in a
unique manner as the direct sum of images of $\cH_0 \otimes \cH_T$ by the
operators $R(k)$ representing elements of $K$ which are {\it not} in $K_0$,
there being one such image for each coset $K_0 k \subseteq K$.
Specifically, we can proceed by choosing a representative element $k$ from
each coset and using the group multiplication rules of $G$ to determine how
a general $R(g)$ acts on the resulting $\cH$.  This process of building up
$R$ from our core UIR is called ``induction'', and there are in fact
several versions of the construction available in the mathematics
literature.  The important point is that $R$ is determined uniquely (up to
equivalence) once the ``core representation'' is given.

\noindent {\bf To summarize:} Given a $K$-orbit $\cO \subseteq \hN$,
let $\sigma$ be the associated projective multiplier and $K_0$ be the
associated little group.  Assume that $\card(\cO) < \infty$.  Then given in
addition a PUIR $T$ of $K_0$ with multiplier $\tau=1/\sigma$, we acquire a
unique $[R]\in {\hat G}$ ($\hat G$ being the set of equivalence classes of
finite dimensional UIR's of $G$).  Moreover, any two $T$'s in the same
similarity class yield equivalent $R$'s.

\noindent {{\underbar {What determines a UIR of G?}}}
\Nobreak

The procedures of analysis and synthesis just described are
inverses of each other.  Hence, taking them together, we have a {\it
bijective correspondence},
\eqn\threef
{
  (\cO, cls(T)) \leftrightarrow [R],
}
where $cls(T)$ is a {\it similarity} class of PUIR's
$T : K_0\rightarrow Aut(\cH_T)$ and $[R]$ is an equivalence class of UIR's
of $G$.  That is: finding all finite dimensional UIR's of $G$ is equivalent
to finding all pairs $(\cO, T)$, where $\cO$ is $K$-orbit in $N$ with
finite cardinality, and $T$ is a collection of operators $T(k), k\in K_0$,
which act irreducibly in a finite dimensional vector space and obey
\eqn\threeg 
{
  T(k_1)T(k_2)=(\sigma (k_1,k_2))^{-1} T(k_1 k_2),
}
$\sigma$ being the projective multiplier associated with the orbit $\cO$.  For
a given $\cO$, two such $T$'s determine the same equivalence class $[R]$
iff they are related by a similarity transformation,
\eqn\threeh
{
  T'(k)=UT(k)U^{-1}  \qquad\quad (U^{*}=U^{-1})
}

We should stress here that a similarity class of PUIR's is not the same
thing as a projective equivalence class.  For the PUIR's $T$ and $T'$ to be
{\it similar}, the equality {\threeh} must hold with a fixed $U$ for all
$k$.  For {\it projective equivalence} an arbitrary $k$-dependent
phase-factor is allowed to intervene.  For projective representations, the
most natural notion of equivalence is projective equivalence.  However, a
projective equivalence class $[T]$ will in general determine a whole {\it
family} of UIR's $[R]$, the elements of the family being related by the
action of the character group of $K_0$ which takes $T\mapsto\chi T$, $\chi$
being the character.  This action leaves the multiplier $\tau$ alone, but
it modifies $[R]$, unless $\chi T$ happens to be similar to $T$.  Thus we
can also think of the UIR's of $G$ as being parameterized (for fixed $\cO$)
by a projective equivalence class $[T]$ of PUIR's $K_0$ together with an
element of the quotient of the (unitary) character group of $K_0$ by the
subgroup of characters that induce similarity transformations on $T$:
$$
       (\cO, [T], [\chi]) \longleftrightarrow [R] .
$$

We should mention here that none of these subtleties involving projective
representations will figure in the part of our subsequent analysis dealing
with the ``particle group'', as in that case, the projective multiplier
$\sigma$ will turn out to be trivial.  However we know of no reason why
nontrivial $\sigma$ would not arise in the analysis of the full MCG (with
the slides playing the part of the normal subgroup $N$, and the particle
subgroup playing the role of the quotient group $G/N=K$).

  In that situation, the full machinery of the present section would be
needed.  Moreover, one would then need to understand certain projective
representations of the little group $K_0\subseteq K$, whence (as $K=\tG$
would itself be a semi-direct product) an analysis of the {\it projective}
representations of semi-direct product groups would likely be pertinent.  A
treatment of that problem can be found in {\mcka}, which indeed furnished
the pattern on which much of the above is based.  However, in {\mcka} the
subtleties involving the character group of $K_0$ do not appear, because
that reference aims only to classify the PUIR's of $G$ up to projective
equivalence, not up to similarity.

It is instructive at this point to apply the preceding analysis to two
special cases which, in the next section, will turn out to correspond to
the two cases where the geons are either all indistinguishable or all
distinguishable, respectively.

\noindent {\bf Case I.}
Let the orbit ${\cal O}$ be such that the associated little group $K_0$ is
all of $K$.  Evidently, this is equivalent to saying that $\cO$ is reduced
to the single equivalence class $[{\Gamma}_0]\in \hN$.   Then $R$ takes the
form
\eqn\rni
{
    R(\dot n)  = {\Gamma}_0(\dot n)  {\otimes} {\bf 1}_{\beta},
}
\eqn\rk
{
    R(\dot k)={S}(\dot k) {\otimes} {T}(\dot k)
}
(where we employ the notation $\dot n$, $\dot k$ to indicate that the
equality holds for {\it arbitrary} elements $\dot k\in K$, $\dot n\in N$).
Recall that ${T}$ is a PUIR of $K$ and ${S}(k)$ is for each $k\in K$ an
intertwining operator between ${\Gamma}_0 (\dot n)$ and ${\Gamma}_0(k\dot
nk^{-1})$.  This UIR of $G$ is obviously finite dimensional if $S$ and $T$
are.

\noindent {\bf Case II.}
This case is the other extreme, where the orbit $\cO$ is such that the
little group associated with any element $[\Gamma_0]\in\O$ is trivial,
i.e., $K_0=\{e\}$.  In this case, $K/K_0=K/\{e\}$ is $K$ itself, $\cO$
contains the maximum possible number $|K|$ of elements of $\hN$ ($|K|$
being the order of $K$), and we have
\eqn\rnii
{
    {\tilde R}=\bigoplus_{i=0}^{|K|-1}{\Gamma}_i ,
}
the multiplicity $\beta$ being one since the little group is trivial and
has only the trivial representation.  Clearly, $|K|$ must be finite in this
case, since we have limited ourselves to finite dimensional representations
$R$.  Finally, we observe that since $T$ is trivial in the present case,
$R(k)$ for $k\in K$ is composed entirely of intertwining maps connecting
the various ${\Gamma}_i$.  Moreover these maps must fit together without
any phase mismatches, since the only possible PUIR of the little group
$K_0=\{e\}$ is the trivial one, for which the multiplier $\sigma \equiv 1$.

Equipped in this manner, we now proceed to the 
physical situation at hand.
\vskip 0.2 true in


\centerline {\bf \S 4. The UIR's that Annihilate the Slides}
\Nobreak
\vskip 0.1 true in

A UIR $R: G \rightarrow Aut(\cH)$ which represents the slides trivially
(annihilates them) is equivalent to a UIR of the quotient group
$G/{\cal{S}}\simeq\tG$, where $\tG$ is the particle group defined earlier.
But $\tG$ is itself the semi-direct product $\tG=G_{int}\semidirect S_N$,
whence it lends itself naturally to the analysis of the preceding section.
This will allow us to find explicitly all the UIR which annihilate the
slides.

Now, in specializing the construction of \S 3 to the present situation we
have $G\rightarrow {\tG}$, $N\rightarrow G_{int}$ and $K\rightarrow S_N$.
Here a crucial simplification occurs because $G_{int}$ is itself a product
of smaller groups, indeed the direct product of $N$ copies of a single
group $G_1$: the MCG of the prime ${\cal P}$, or more precisely of
$\cP\#\R^3$.  (Recall we have confined ourselves to 3-manifolds of the form
$\R^3\#\cP\#\cP\#\cdots\#\cP$).  Since $G_{int}$ is a direct product, its
UIR's are the tensor product representations
${\Delta}_1\xsq{\Delta}_2....\xsq{\Delta}_N$, where each ${\Delta}_i$ is
some UIR of the internal diffeo group of a single prime.  From this it
follows that the operators $T(k)$ of \S 3 can be chosen {\it canonically},
and they will then manifestly furnish an {\it ordinary} representation of
$K_0$ (i.e. one with $\sigma \equiv 1$).  This follows from the fact that
$T(k)$ just acts by permuting the factors of a tensor product, and such a
permutation can always be implemented canonically -- by an operator we will
call $\cE$.  For example in $V\otimes V$ the exchange $\cE$ taking
$\psi_1\otimes\psi_2\rightarrow\psi_2\otimes\psi_1$ is the canonical
choice, and cannot be mistaken for
$\psi_1\otimes\psi_2\rightarrow-\psi_2\otimes\psi_1$.

This uniqueness of the $\cE$'s is important for another reason.  Not only
does it save us from having to consider projective representations in
constructing ${\hat\tG}$ (because
$\sigma\ideq{1}\Rightarrow\tau=1/\sigma\ideq1$), but with respect to
physical interpretation, this absence of phase-ambiguity will allow us to
draw a meaningful distinction between the UIR's describing bosonic geons
and those describing fermionic geons.

At this point we observe how the quantum behavior of otherwise identical
primes can in fact render them distinguishable.  We started off with $N$
classically identical primes whose internal groups $G_i$ are therefore all
isomorphic.  However these internal groups might all be represented by
inequivalent UIR's  $\Delta_i$.  For example, if the internal group was
generated by the $2\pi$ rotation, i.e., $G_i=\{e, {\cal R}_{2\pi}\} \simeq
\Z_2$, then ${\cal R}_{2\pi}$ could be represented either spinorially or
tensorially {\fsor}.  Thus a generic UIR of the diffeo group of a set of
such classically identical primes has both spinorial and tensorial geons,
which accordingly belong to distinct species of particles.  This then, is an
example of what we referred to earlier as the quantum breaking of
indistinguishability.  (At the same time, one is led to the puzzle of what
role the exchange operator plays when the geons are rendered
distinguishable.  We will comment on this later in the section).

To illustrate things more clearly, we consider the two simplest non-trivial
cases, in which the manifold is made up of two and three identical primes
respectively.  One can readily infer from these examples what the structure
of the UIR of ${\tilde G}$ is for an arbitrary number of identical primes.

\bigskip

\noindent { {\bf Two identical Primes ($N=2$)}}

Consider the situation where the three manifold $\M$ is $R^3\#\cP\#\cP$,
where ${\cP}$ is an irreducible prime.  Then, $G_{int}=G_1\times G_1$ where
the isomorphic internal groups $G_1$ and $G_2$ of the two primes have both
been identified to $G_1$ (the particular choice of identification following
from our choice of presentation of $\M$).  The permutation subgroup is then
$S_2\simeq\Z_2$ and correspondingly, $\tG=(G_1\times{}G_1)\semidirect\Z_2$,
with $\Z_2$ acting on $G_1\times G_1$ by exchange of the $G_1$ factors.
Let $R$ be a UIR of ${\tilde G}$ and ${\tilde R}$ its restriction to
$G_{int}$.  We must first find the orbits of the action of $S_2$ on the
space $\widehat{G_{int}}$ of finite dimensional UIR's of $G_{int}$.  These
orbits can be classified into two types, namely (a) orbits corresponding to
identical geons and (b) orbits corresponding to distinguishable geons.

\noindent{\underbar {Type (a) orbit}}
\Nobreak
Here the fiducial element of our orbit is a UIR,
${\Gamma}\in{\hat{G}}_{int}$ which has the form
${\Gamma}=\Delta\xsq\Delta$, where $\Delta$ is a UIR of $G_1$.  (We omit
the subscript $0$ from ${\Gamma}$ since it is not needed in this
case).  The two primes therefore represent quantum mechanically identical
geons.  How does $S_2$ act on such a UIR?  The only non trivial element of
$S_2$ is the exchange $E$, and we observe that its action on ${\Gamma}$ is,
\eqn\Ga
{
  {\Gamma} (g_1, g_2)= \Delta(g_1)\otimes \Delta(g_2) \mapsto
  {\Gamma}'(g_1, g_2)={\Gamma}(g_2,g_1)=\Delta (g_2)\otimes\Delta (g_1).
} 
But these two UIR's are equivalent to each other via the {\it canonical
exchange} $\cE(E)$ which just reverses the orders of the factors in the
carrier space $\cH_{\Delta} \otimes \cH_{\Delta}$ of $\Delta \xsq\Delta$.
Hence, the little group associated with the UIR $\Gamma$ is $S_2$ itself.
This is an instance of case I of \S 3, so
\eqn\Gna 
{
 {\tilde R}=(\Delta  \xsq \Delta){\otimes} {\bf 1}_{\beta},
}
with $\beta $ being the dimension of some UIR of $S_2$. Now section 3 tells
us that $\tR(k)=S(k) \otimes {T}(k)$, where $k\in S_2$.  But $S$ is just
the (ordinary) representation of $S_2= \{1,E\}$ given by $S(1)=1$ and
$S(E)=\cE$, whence (since $S$ has trivial projective multiplier) $T$ must
be an ordinary UIR of $S_2$.  Hence $T$ is either trivial or is the
1-dimensional representation given by $T(E)= {-1}$ (with of course
$T(1)=1$).  Thus the full representation $R$ of $\tG$ is given in this case
by
$$
               \cH =   \cH_\Delta \otimes \cH_\Delta
$$
\eqn\foura
{
  (g_1,g_2)\mapsto \Delta(g_1)\otimes \Delta(g_2), \ \ g_i \in G_1 ,
}
\eqn\fourb
{
     E \mapsto \pm \cE,
}
where the two signs correspond to the two possibilities for $T$.


The physical interpretation for these two UIR's is evident.  The geons are
not merely identical classically, i.e. diffeomorphic, but are identical
quantum mechanically; their internal groups are represented by the same
UIR.  Thus the exchange operator is physically meaningful, even at the
quantum level, and this is reflected in the fact that $T$ can be a
non-trivial representation of the permutation group, giving rise to the two
different possible statistics, bosonic and fermionic.  At this stage, one
also sees the well known spin statistics violation emerging because of the
possibility of arbitrary combinations of internal representations with
statistics.  In other words, given that the internal group is tensorially
(spinorially) represented, i.e., $R_{2\pi}$ is represented trivially
(non-trivially), there is nothing to prevent the permutations from being
fermionic (bosonic).  Indeed we see that given {\it any} two identical
primes, there exists at least one quantum sector for which the spin
statistics correlation is violated.

\bigskip
\noindent {\underbar {Type (b) orbit}}
\Nobreak
The other form that the fiducial UIR of the internal group $G_{int}$ can
take is ${\Gamma}_0=\Delta_a \xsq \Delta_b $ where $\Delta_a$ and
$\Delta_b$ are {\it inequivalent} UIR's of the internal group of a single
prime.  Now the exchange $E$ has the action, 
\eqn\Gb
{
  \Delta_a(g_1) \otimes \Delta_b (g_2) \mapsto \Delta_a(g_2) \otimes
  \Delta_b(g_1) \simeq \Delta_b(g_1)\otimes \Delta_a(g_2).
}
That is, ${\Gamma}_0=\Delta_a \xsq \Delta_b$ is taken into the inequivalent
representation ${\Gamma}_1=\Delta_b \xsq \Delta_a$.  Thus the little group
associated with ${\Gamma}_0$ is trivial and one is in case II of \S 3 with
$\cO=\{\Delta_a\otimes\Delta_b,\Delta_b\otimes\Delta_a\}$.  Therefore
(since the little group is trivial),
\eqn\Gnb
{
  {\tilde R}=\Delta_a\xsq\Delta_b  \ \oplus \  \Delta_b\xsq\Delta_a.
}
Using a block matrix notation, we can write
$$ 
{
{\tilde R}=\pmatrix { \Delta_a \xsq \Delta_b &   0                     \cr
                        0                    & \Delta_b \xsq \Delta_a  \cr }
}
$$
and
$$
{
  {R}(E)=\pmatrix { 0      & {\cal E}^{-1}\cr
                  {\cal E} &  0           \cr }
}
$$ 
where now ${\cal E}:\cH_a\otimes\cH_b\rightarrow\cH_b\otimes\cH_a$ is the
operator which takes $\psi_a\otimes \psi_b \mapsto \psi_b \otimes\psi_a$
($\cH_a$ being the carrier space of $\Delta_a$ and $\cH_b$ that of
$\Delta_b$).  The form of this UIR of ${\tilde G}$ is therefore completely
determined.

In this case, we have a situation in which the geons are rendered
distinguishable by the fact that their internal groups are represented
inequivalently, and so the notion of statistics loses its meaning for this
UIR because the operator $R(E)$ no longer has the significance of an
exchange of identical objects.  Consider for example a wave function in
Hilbert space which is peaked at a point of $Q$ for which the primes have
definite spatial locations with respect to a flat ambient geometry.  For
such states the representation $R$ gives direct information on the results
of physical operations on the geons: exchange via parallel transport in the
case of $E$, and local rotation or other ``internal deformations'' in the
case of $G_1$ and $G_2$ (see {\AmalfiBal}, {\amalfi}).  Let $\psi_{ab}$ be
one such wave-function for which the geon at location 1 is ``of species
$a$'' (i.e., carries the representation $\Delta_a$) and that at location 2
is ``of species $b$''; and let $\psi_{ba}=R(E)\psi_{ab}$ be the exchanged
state.  Clearly $\psi_{ba}$ is physically distinct from $\psi_{ab}$ (an
example of ``quantum multiplicity''), since for it the species of the geons
at locations 1 and 2 have been swapped.  Now, nothing  stops us from
forming the even and odd eigenvectors $\psi_{ab} \pm \psi_{ba}$ of $R(E)$,
but the sign which occurs in these combinations has nothing to do with bose
or fermi statistics, because the particles being ``exchanged'' are not
identical.  (See the discussion in \S 6 below, however.)  It is exactly as
if a proton and a neutron were present and $R(E)$ were the operator which
exchanged their positions.  That operator could not be used to determine
nucleon statistics, and indeed it is entirely a matter of convention
whether ``the nucleon'' is a boson or a fermion, as long as one has only a
single $p-n$ system to refer to.

\bigskip
\noindent { {\bf Three Identical Primes ($N=3$)}}

Let $\M=R^3 \# \cP \#\cP\#\cP$.  This case is interesting
because it is the simplest one in which non-abelian statistics
(parastatistics) can occur.  Moreover there are orbits in $\hat{G_{int}}$
for which the little group is neither the whole permutation group $S_3$,
nor trivial, and hence we need to use a greater portion of the apparatus
developed in \S 3.  The orbits in this case fall into three classes, (a)
all three geons are identical, (b) all three geons are distinct, and (c)
two geons are identical but distinct from the third.  The analysis for
orbit types (a) and (b) is similar to the what we saw in the $N=2$ case
above, except that in case (a) the little group is $S_3$ rather than $S_2$,
and therefore its UIR $T$ can be non-abelian as well as bosonic or
fermionic
(see equation (29) below).  Thus not only is there a possibility of the
violation of the usual spin-statistics correlation but also of the
occurrence of parastatistics.  Indeed, sectors manifesting bose, fermi and
parastatistics exist for $every$ irreducible prime $\cP$ and there is
nothing in the present analysis to rule out their appearance physically.
We proceed to case (c).

 The fiducial UIR of the internal group for an orbit of type (c) has the
form ${\Gamma}_0=\Delta_a\xsq \Delta_a \xsq \Delta_b$.  The little group
associated with this UIR is the $\Z_2$ subgroup $K_0=\{ e, P_{12}\}$, where
by $P_{ij}$ we mean the exchange of the primes $i$ and $j$.  Forming
right cosets of $K_0$ in $S_3$ yields the partition,
\eqn\perm
{
S_3=K_0\amalg K_0  P_{13} \amalg K_0  P_{23},
}
so $\Gamma_0$ engenders an orbit of three elements given by the action of
$P_{13}$ and $P_{23}$ on it.  Therefore
\eqn\Gc
{
   {\tilde R} = 
	  \Delta_a \xsq \Delta_a \xsq \Delta_b  \ \oplus\ 
          \Delta_b \xsq \Delta_a \xsq \Delta_a  \ \oplus\ 
          \Delta_a \xsq \Delta_b \xsq \Delta_a ,
}
since the UIR's of $\Z_2$ are one dimensional, whence the multiplicity
$\beta$ is one.  Since the little subgroup $K_0\subseteq{}S_3$ governs the
statistics, and since $K_0=\Z_2$ has only a bosonic and a fermionic
representation, we find again that the pair of identical geons can obey
either bose or fermi statistics, again irrespective of whether their
internal groups are represented spinorially or tensorially.

\bigskip
\noindent {\bf The general case}

What can we infer from the above analysis for the general case of $N$
identical primes?  If the fiducial UIR of the internal group $G_{int}$ is
\eqn\Ggen
{
 {\Gamma}_0= {\underbrace {\Delta_1 \xsq \cdots \xsq \Delta_1}_{N_1}}
  \xsq {\underbrace{\Delta_2 \xsq \cdots\xsq\Delta_2}_{N_2}}\cdots
  \xsq {\underbrace {\Delta_m\xsq\cdots\xsq \Delta_m}_{N_m}}
}
where $N_1 + N_2 + \cdots + N_m=N$, then the little subgroup of $S_N$ 
associated with it will be 
$K_0=S_{N_1}\times S_{N_2}\times\cdots\times  S_{N_m}$.  To obtain a
specific UIR $R$ of $\tG$ we must then specify a UIR $T$ of $K_0$, or
equivalently a UIR for each of its factor groups,
$S_{N_1},S_{N_2},....,S_{N_m}$.  {\it Thus the general UIR of $\tG$ in the
case of $N$ identical primes is given by the following}:

\item{$\bullet$} a choice of up to $N$ inequivalent UIR's
  $\Delta_1,\Delta_2,...,\Delta_m$ of 
  $G_1 = \pi_0 (\Diff^{\infty}(\cP\# {\R}^3))$

\item{$\bullet$} For each $\Delta_i$ a multiplicity N$_i$, such that
         $\Sigma N_i=N$.

\item{$\bullet$} For each $\Delta_i$ a UIR $T_i$ of $S_{N_i}$, the permutation
   group on $N_i$ elements (i.e. a Young tableau with $N_i$ boxes).
 
\noindent
Physically the resulting $R$ describes a quantum sector in which there are
present $N_i$ geons of type $\Delta_i$ bearing the statistics defined by
$T_i$.  We see in particular that all possible parastatistics for a given
$N$ can occur, and also that spin-statistics violating sectors will always
exist for any set of identical irreducible primes.

The general case of an arbitrary (finite) collection of primes is an
obvious generalization of the above, and need not be treated explicitly
here.  However we would like to stress that the quantum sectors which arise
in the general case are neither more nor less than what would be suggested
by the interpretation of geons as particles: a certain number of geons of
each species is present and each species can have its own statistics.  If
there are any surprises or subtleties at this stage, they are associated
only with the fact that the underlying prime manifold is not enough to
determine a geon's species, but a UIR $\Delta $ of the prime's MCG is
needed as well (``quantum breaking of indistinguishability'').  In addition
the complete decoupling of spin from statistics could seem surprising, but
that is already an old story.

In addition to a classification of the UIR's it might be useful to have
their full forms presented explicitly.  Such a presentation in the general
case seems too cumbersome to be illuminating, but the following two extreme
cases are representative.  They correspond to cases I and II of \S 3.

\noindent {\bf {All geons identical: case ${\rm I}'$}}

Here there is a single UIR $\Delta =\Delta_1$ of $G_1$, so $m=1$, $N_1=N$ and
$K_0=S_N$, the full permutation group on $N$ elements.
For $(g_1, g_2...g_N) \in G_{int}=G_1\times G_1 \times....\times G_1$, we
have from {\rni}
\eqn\Gngen
{
  {R}((g_1, g_2,...g_N)) = 
 [\Delta(g_1) \otimes .....\otimes \Delta(g_N) ] {\otimes} {\bf 1}_{\beta},  
}
where ${\bf 1}_{\beta}$ is the identity operator in $\cC^{\beta}$,
while for $p\in S_N$, we have from {\rk}
\eqn\Gpgen
{
  {R}(p)={\cE}(p)\otimes{T}(p),
}
where ${T}$ is an arbitrary UIR of $S_N$ and $\cE(p)$ is the canonical
permutation of the factors of
$\cH_{\Delta}\otimes\cH_{\Delta}\otimes\cdots\otimes \cH_{\Delta}$,
$\cH_{\Delta}$ being the carrier space of $\Delta$.

\noindent {\bf  All geons distinct: case ${\rm II}'$}

Here all the  $\Delta_i$ are distinct, so $N_1=N_2=\cdots=N_N=1$, $K_0=\{e\}$
and 
\eqn\Gngenii
{
  {\tilde R}=[\Delta_1 \xsq \Delta_2 \cdots\xsq \Delta_N] \oplus [\Delta_2\xsq
  \Delta_1 \cdots\xsq\Delta_N]\oplus \cdots\cdots
}
as in {\rnii}, while
\eqn\Gpgenii
{
R(p)={\cal E}(p).
}
As in the analogous cases above for 2 and 3 primes, no notion of
statistics enters here since the geons are all distinct. 

\vskip 0.2 true in

\centerline
  {\bf \S 5. The General UIR of $G$ and Two Special Cases}
\Nobreak
\vskip 0.1 true in

In the preceding section we have seen that the structures which appeared in
the general analysis of \S 3 have a direct physical meaning when $G$ is
the particle group, $\tG=G_{int}\semidirect  S_N$.  As applied to $\tG$,
that 
analysis told us that the general (finite dimensional) UIR of $\tG$ can be
obtained from a UIR of $G_{int}$ together with a UIR of the corresponding
little group $K_0\subseteq S_N$.  Physically the UIR of $G_{int}$ (or rather
its $S_N$-orbit in $\widehat{G_{int}}$)
determines the {\it  species} of the geons
which are present, while the UIR of $K_0$ determines their {\it 
statistics}.

The above analysis covers all those UIR's of $G=\pi_0(\Diff^{\infty}(\M))$
for which the slide subgroup $\cal S$ is represented trivially.  If we drop
this restriction, then we obtain a more complicated situation in which
different decompositions of $G$ are possible.  Now, we have a nested triad
of normal subgroups,
\eqn\fivea
{
 {\cal S} \subseteq {\cal S}G_{int} 
 \subseteq {\cal S}G_{int}S_N={\cal S}\tG=G,
}
and the analysis of \S 3 can be applied in stages.  One possibility would
be to build up $G$ as
\eqn\fiveb
{
  G=({\cal S}\semidirect G_{int})\semidirect S_N,
}
which would lead to first finding the UIR's of ${\cal S}$, then those of
${\cal S}\semidirect G_{int}$, and finally those of $G$.  If in the process
a proper subgroup of $G_{int}$ emerged as a little group, we could speak of
a ``quantum breaking of internal symmetry''.  Similarly, if a proper
subgroup of $S_N$ appeared as a little group at the next stage, we could
speak of a ``quantum breaking of indistinguishability'' as before.  A
related scheme would analyze $G$ as
\eqn\fivec
{
G={\cal S}\semidirect {\tG}={\cal S}\semidirect (G_{int}\semidirect S_N)
}
in which case we would obtain a single little group as a subgroup of $\tG
=G_{int} \semidirect  S_N$ rather than a pair of little groups in $G_{int}$
and $S_N$ 
separately.  The resulting description of a given $R\in \hG$ would not
appear to differ significantly from that tied to the  
decomposition {\fiveb}.

Once one had obtained a UIR $R$ of $G$, it could also prove informative to
decompose it into a sum of UIR's of $\tG=G_{int}S_N$, rather than of ${\cal
S}$ or ${\cal S}G_{int}$.  Although such a decomposition is not of the type
studied in \S 3 (because $\tG$ is not normal in $G$), it could often be the
most appropriate
 physically.  Decomposing $R$ in this way would lead to a
description in which the slides did not break internal symmetries or geon
distinguishability, but rather {\it induced transitions} between different
types of geons (inelastic scattering).  We suspect that the different
languages associated with these different ways of decomposing $R$ might all be
useful depending on the physical situation under study.  For example the
decomposition corresponding to ${\cal S}\subseteq {\cal S}G_{int}\subseteq
G$ might be suitable for a hot dense ``gas'' of geons, whereas that
corresponding to $\tG \subseteq G$ might be best for talking about
geon-geon scattering.

In any case, none of these contemplated analyzes can be completed without
an understanding of the space $\hcS$ of UIR's of $\cS$, or at least some
understanding of what types of little groups are induced by the action of
$\tG$ on $\hcS$.  In addressing this question, one has available 
(from \S 2) the knowledge of how the permutations and the internal
diffeomorphisms act on a  generating set of slides, and in principle
this is all we would need to know in determining the $G_{int}$ or $\tG$
orbits in $\hcS$.


Let us remark here that the subgroup of the diffeos consisting of only the
$2\pi$ rotations, which we shall call ${\cal R}_{2\pi}$, in fact commutes
with all diffeos and hence will always be in (the center of) the little
group of any element of ${\hat {\cal S}}$

Unfortunately, an analysis of $\hat G$ as complete as that of
$\hat{\tilde{G}}$ in \S 4 seems out of reach.  To start with, the UIR's of
$\cS$ are much harder to characterize than were those of $G_{int}$, an
important difference being that $G_{int}$ is a {\it direct product} whereas
$\cS$ in general involves {\it free products}.  Similarly, the action of
$\tG$ on $\cS$ is more complicated than was that of $S_N$ on $G_{int}$.
Moreover, the seeming lack of a canonical choice for little group operators
of the type $S(g_{int})$ or $S(p)$ means that nontrivial projective
multipliers can be expected to come into play when the slides are
nontrivially represented.  For all these reasons we are unable to offer
complete results in the general case.  Nevertheless, there are still
special situations in which the analysis can be carried through, and
already in these special cases we will encounter interestingly novel types
of particle behavior.  In the remainder of this section we consider two
such situations.

\noindent 
\underbar{Case ($\A$)}
Let ${\cal P}$ be a prime whose internal group is trivial
(${\cal{P}}\simeq\R{{P}}^3$).  Then $G_{int}=\{e\}$ and we have simply
$G={\cal S} \semidirect S_N$.  Further, let us select an orbit $\cO$ of the
$S_N$ action on ${\hat {\cal S}}$ whose associated little group is also
trivial.  We are then in case II of \S 3 (with
$N\rightarrow\cS,K\rightarrow S_N$); $\cO$ is essentially a copy of $S_N$
itself, and we acquire from $\cO$ a unique UIR $R$ of $G$, for which
\eqn\shani
{
  \tR\equiv R|\cS 
     = {\Gamma}_0 \oplus {\Gamma}_1 \oplus \cdots \oplus {\Gamma}_{m-1} 
}
where $m=N!$ is the order of $S_N$ and the ${\Gamma}_i$ are the elements of
$\cO$.  Since the little subgroup of $S_N$ is trivial, the UIR $T$ is also
trivial, and the possibility of different types of statistics does not
arise.

In \S 4 we interpreted a similar situation to mean that the geons were
quantum mechanically inequivalent because their internal groups were
represented differently.  However in the present case the internal groups
are all trivial and hence represented identically.  What can the
inequivalence of the UIR's ${\Gamma}_{j}$ mean physically in this case?
Consider an element of $S_N$, say the exchange $p_{ij}$ of the $i^{th}$
with the $j^{th}$ prime.  Since the little group associated with the
representation ${\Gamma}_0$ is trivial,
${\Gamma}_0\mapsto{\Gamma}_k\not\simeq{\Gamma}_0$ under the action of
$p_{ij}$.  Now, $p_{ij}$ will commute with all slides that do not involve
the $i^{th}$ or the $j^{th}$ prime.  Thus if ${\Gamma}_k$ is inequivalent
to ${\Gamma}_0$, it can only mean, for example, that the slide of the
$i^{th}$ geon through the $j^{th}$ is not the same, physically, as that of
the $j^{th}$ geon through the $i^{th}$.  We encounter here a quantum effect
unlike that in any other system to our knowledge, namely that slides
(essentially a type of collision) render distinguishable particles which
cannot be told apart if examined separately.  In other words, if one merely
examined the individual properties of these geons, they would seem to be
identical, but if they were made to interact appropriately with each other,
one would find that they were in fact different.  We will encounter a
concrete example of this effect in \S 6.

\noindent 
\underbar{Case ($\B$)}
The situation just discussed manifested a ``quantum breaking'' of
permutation symmetry.  A different possibility is that the permutations
remain intact while the {\it internal} symmetry is broken.  To see how this
might occur, consider the decomposition $G={\cal S} \semidirect \tG$ (with
$K:=\tG=G_{int}\semidirect S_N$) and suppose that ${\Gamma}_0$ is a UIR
of ${\cal S}$ for which the little subgroup $K_0\subseteq{\tilde G}$ is
reduced to $K_0=S_N$.  (This implies that ${\cal R}_{2\pi}$ must be trivial
as remarked earlier).  Since $\tG=G_{int}\semidirect S_N$, the coset space
$K_0\backslash K = S_N\backslash\tG$ can be identified with $G_{int}$ and
we have
\eqn\quot
{
{\tilde G}=\amalg_{\alpha}S_N b_{\alpha}, 
}
where the $b_{\alpha}$ run through $G_{int}$ as $\alpha$ runs from $0$ to
$m-1$, with $m=|G_{int}|$.  (Note however that $ S_N\backslash\tG $ is not a
group since $S_N$ is not normal in $\tG$).  The ``core'' representation
thus takes the form
\eqn\another
{
  S\otimes T
}
where $S$ is a PUIR acting on the carrier space of ${\Gamma}_0$ and $T$ is
a PUIR of $S_N$ of dimension ${\beta}$ which we can interpret as
determining the statistics of the geons.  [We know of no reason why $S$,
and therefore $T$, would not be properly projective in general.  This would
produce a new kind of ``generalized statistics'' determined by a {\it
projective} representation of the permutation group rather than an ordinary
one.  (Non-trivial projective representations of $S_N$ exist for $N\geq
4$.)].  It follows that
\eqn\sb
{
 {\tR}=[{\Gamma}_0 \oplus \cdots  {\Gamma}_{m-1}]{\otimes} {\bf 1}_{\beta}
}
and the dimension of our UIR is $m\beta(\dim\Gamma_0)$.  (So we need
$|G_{int}|<\infty$, in order to comply with our standing restriction to
finite dimensional representations).

Since the little group is all of $S_N$ in this type of example, the geons
must all be identical.  What does it mean, however, that $G_{int}$ is not
part of the little group?  Classically an internal diffeo corresponds to an
internal symmetry of a prime $\cal P$.  If such a diffeo takes a generator
$\phi\in \pi_1({\cal P})$ to a generator $\phi '$, then this means that
$\phi$ and $\phi '$ are in a sense equivalent.  On quantization, however,
if the UIR's ${\Gamma}_i$ are {\it not} left invariant under the action of
$G_{int}$ (as in {\sb}), then this must mean that the slides of some other
prime along the two generators are inequivalent.  In other words, the
slides have rendered the geons ``asymmetric''!  Again we observe that
merely examining an isolated geon is not enough to determine whether it is
symmetric.  One requires ``collisions'' with the other geons in order to
determine whether the classical internal symmetry has been broken or not.
This type of symmetry breaking also appears to be novel.

In the next section we will construct examples of case $\A$ where $\M$ is
the connected sum of $\R^3$ with $N$ copies of $\R{}P^3$, and where
$\Gamma_0$ is an abelian UIR of $\cS$.  One can also construct examples of
case $\B$ in a similar manner, but since we lack a systematic analysis, we
will not present any explicit examples of case $\B$.


\vskip 0.2 true in
\centerline {\bf \S 6. The Example of $\R{P}^3$ Geons}
\Nobreak
\vskip 0.1 true in

Let the manifold $\M={\R}^3 \# {\underbrace {\R{P}^3\cdots\#\R{P}^3}_{N}}$.
(The connected sum of ${\R}^3$ with $N$ ${\R}P^3$ primes can be visualized
as follows: take ${\R}^3$ and remove $N$ three balls ${\D}^3$ from it;
then perform an antipodal identification for each of the resulting $S^2$
boundaries.)  The internal diffeo group for an ${\R}{P}^3$ prime is
trivial, and therefore the MCG of $\M$ is ${\cal S}\semidirect{}S_N$.
Moreover, $\pi_1(\R{P}^3)={\Z}_2\Rightarrow {s}^2=e$ for any slide $s$.
The slide subgroup therefore has $N(N-1)$ generators $s_{ij}$ satisfying
$s_{ij}^2=e$ and satisfying the Fuks-Rabinovitch commutation relations
given in \S 2.  Notice that the latter become entirely trivial for any
1-dimensional representation of the slides.

\noindent {\underbar {\bf A pair of $\R{P}^3$ geons}}

The mapping class group $G=\pi_0(\Diff(\R^3 \# \R{P}^3\# \R{P}^3))$ is the
semidirect product $G={\cal S}\semidirect S_2$, where the nontrivial
element $E\in{S_2}$ interchanges the two primes, and the slide subgroup,
$\cS=\Z_2*\Z_2$, is generated by the slide $s_1$ of the first prime through
the second together with the slide $s_2$ of the second prime through the
first {\fuksrab} {\mcc}.  Clearly $E$ acts on the normal subgroup $\cS$ by
exchanging its two $\Z_2$ factors.  A set of generators and relations for
$G$ can then be given as follows.
\eqn\PresOne
 {  G = < s_1, s_2, E : s_1^2=s_2^2=E^2=1, Es_1=s_2E >   }
Notice that not only $\cS$ but $G$ itself is isomorphic to $\Z_2*\Z_2$, as
can be seen from the following alternative presentation.
\eqn\PresTwo
{
  G = < x, y : x^2=y^2=1 > ,
}
where $x=s_1$ and $y=E$.  (Of course an infinite group can be isomorphic
to a proper subgroup without contradiction.)

The presentation {\PresTwo} reduces the problem of finding the UIR's of $G$
to that of finding the UIR's of $\Z_2*\Z_2$.  This could be done directly
(and has been in {\amalfi}), however for present purposes it is more
instructive to build up the UIR's of $G$ from those of $\cS$ and $S_2$ by
applying the method of \S 3 with $N\to\cS$, $K\to S_2$.  In this way,
we will be able to identify the appropriate little groups in each case,
enabling us to draw conclusions about geon indistinguishability and
statistics.

Following the pattern of \S 3, our first step is to determine the
UIR's of $\cS$ and the orbits $\O$ induced in $\cS$ by the action of $S_2$.
The UIR's of $\cS\isom\Z_2*\Z_2$ may themselves be found by the method of
\S 3, taking for the normal subgroup $N'\subseteq \Z_2 * \Z_2$ the
cyclic subgroup $\{(xy)^n | n\in \Z \}\isom\Z$.  With this identification,
$N'$ is abelian and its character group $\widehat{N'}$ contains elements
$\Gamma_\theta$ parameterized by an angle $\theta\in\R\,\mod\,2\pi$
(i.e. $\Gamma_\theta(n)=q^n$ with $q=e^{2\pi{i}\theta}$ ).  Under the
action of $\Z_2 = \Z_2{*}\Z_2/N'$, there are a pair of singleton orbits
$\O=\{\Gamma_0\}$, $\O=\{\Gamma_\pi\}$ and an open interval's worth of
2-element orbits $\O=\{\Gamma_\theta,\Gamma_{-\theta}\}$, where
$0<\theta,\pi$.  The corresponding UIR's are readily found, and are listed,
in a symmetrically chosen basis, in Table 1 (wherein the parameters $\tau$
and $\gamma$ are respectively $\cos\theta/2$ and $\sin\theta/2$, and the
first four UIR's correspond to $\theta=0$ and $\theta=\pi$).  Having now
determined the space $\hat{\cS}$, our next step is to fiber it into orbits
$\O$ under the action of $S_2=G/\cS$.

Under exchange of the slides $s_1$ and $s_2$, the UIR's in table 1
fall into orbits of four types.  The first two UIR's are evidently
fixed points, and hence each constitutes an orbit in itself.
Similarly, the third and fourth UIR's combine to form a single orbit.
Finally, each 2-dimensional UIR is also a fixed point, since the
exchange can be implemented unitarily by the operator $S(E) :=
\pmatrix{1 & {\, 0}\cr 0 & {-1}}$.  Taking these cases in order, we get the
following decomposition of $\hat{\cS}$ into disjoint orbits (using the
names given in table 1):
\eqn\orb
{
  \hat{\cS} = \{\Gamma_{++}\} \cup
             \{\Gamma_{--}\} \cup
             \{ \Gamma_{+-}, \Gamma_{-+} \} \cup          
             \bigcup\limits_{0<\tau<1} \{ \Gamma_\tau \}
}
Evidently the associated little groups $K_0\subseteq S_2=\Z_2$  are
\eqn\lgr
{
  K_0=\Z_2,  \quad  K_0=\Z_2, \quad  K_0=\{e\}, \quad   K_0=\Z_2
}
and for the representations $S$ of $K_0$, we may choose 
\eqn\slgr 
{
 S(E)=1 ,\quad 
 S(E)=1 ,\quad
 S  = 1 ,\quad 
 S(E)=\pmatrix{1 & { 0}\cr 0 & {-1}} 
}
(Here the first three choices of $S(E)$ are trivial of course, since all
the UIR's involved are 1-dimensional).  Given these orbits, little groups
and operators $S(E)$, we can proceed to construct the UIR's of
$G=\cS\semidirect S_2$.

\bigskip\bigskip

\vbox{
\centerline {\bf Table 1: UIR's of the slide subgroup $\Z_2 * \Z_2$}
\centerline{The  final column shows the ``Casimir invariant''
              $\half\Gamma(s_1 s_2 + s_2 s_1)$ }
$$
\vbox{                       
\halign{ # \hfil & \quad 
         # \hfil & \quad 
         # \hfil & \quad 
         # \hfil & \quad 
         # 
        \cr
\noalign{\hrule}
\noalign{\smallskip}
\noalign{\hrule}
\noalign{\medskip}
  Dim            &  
  $\Gamma(s_1)$  &  
  $\Gamma(s_2)$  & 
  Name           &
  Invariant      \cr
\noalign{\medskip}
\noalign{\hrule}
\noalign{\medskip}
 {  1}  & {  1}   & {  1}  & $\Gamma_{++}$ & $+1$ \cr     
 {  1}  & {$-1$}  & {$-1$} & $\Gamma_{--}$ & $+1$ \cr     
 {  1}  & {  1}   & {$-1$} & $\Gamma_{+-}$ & $-1$ \cr     
 {  1}  & {$-1$}  & {  1}  & $\Gamma_{-+}$ & $-1$ \cr     
        &         &        &                      \cr     
 {  2}  &			                          
  {$\pmatrix{\tau & \gamma \cr
           \gamma &-\tau \cr}$}     & 
              
  {$\pmatrix{\tau & - {\gamma} \cr
       -{ \gamma} &-\tau \cr}$}     &  
   $\Gamma_\tau$                    &
   $2\tau^2-1$                      \cr 

        &       &     &                 \cr          
    { } &  $\tau\in(0,1)$ &  $\gamma={\sqrt{1-{\tau}^2}}>0$ & { } \cr
\noalign{\medskip}
\noalign{\hrule}
\noalign{\smallskip}
\noalign{\hrule}
}}$$}

\bigskip
\noindent 
{\bf { The orbits $\{ \Gamma_{++} \}$ and $\{ \Gamma_{--} \}$}}

Both of these cases are essentially trivial since $\Gamma$ is 1-dimensional
and the little group $K_0$ of {\lgr} is $\Z_2$, which has only the two
1-dimensional UIR's $T(e)=1$, $T(E)=\pm{1}$.  (Projective multipliers
obviously do not arise in this situation, since $\sigma\ideq{}1$ for the
trivial representations $S$ of \slgr.)  Since $\O$ consists of a single
point, our ``core'' representation {\threee} is already all of $R$, and we
obtain the $2\times 2=4$ abelian UIR's of $G$, which are listed in Table 2.
(The first and third come from $\Gamma_{++}$, the second and fourth from
$\Gamma_{--}$.)

Since the little group in these two cases is $S_2$, we may interpret
these UIR's as describing identical geons with statistics given by $T(E)$.
This obviously makes physical sense since $R(E)$ is a simple sign in each
case.  The first two representations in Table 2 describe bosonic geons,
and the second two describe fermionic ones.

\noindent 
{\bf  {The orbit $\{ \Gamma_{+-} ,\,  \Gamma_{-+} \}$}}

Here the little group is trivial and so (see eq. {\rnii} above)
$$
     \tilde R = \Gamma_{+-}  \oplus \Gamma_{-+} ,
$$
or in a matrix notation,
\eqn\sone
{
   R(s_1)=\pmatrix {1 & { 0} \cr 
                    0 & {-1} \cr},
}
\eqn\stwo
{
   R(s_2)=\pmatrix {{-1} & 0 \cr 
                     { 0} & 1 \cr}.
}
The exchange operator then transforms {\sone} into {\stwo}:
\eqn\enotid
{
   R(E)=\pmatrix{ 0 & 1  \cr
                  1 & 0  \cr}.
}
This non-abelian UIR of $G$ furnishes the fifth row of table 2.

\bigskip
\vbox{
\centerline 
  {\bf Table 2: UIR's of the mapping class group $G$ for two ${\R}P^3$ geons} 
$$
\vbox{                    
\halign{# \hfil & \quad 
        # \hfil & \quad 
        # \hfil & \quad 
        # \hfil & \quad 
        # \hfil & \quad 
        #
\cr
\noalign{\hrule}
\noalign{\smallskip}
\noalign{\hrule}
\noalign{\medskip}
{Dim} &  { } & $R(s_1)$  &  $R(s_2)$ & $R(E)$ \cr  
\noalign{\medskip}
\noalign{\hrule}
\noalign{\medskip}
 {  1}  & {  }  & {  1} & {  1} & {  1} \cr
 {  1}  & {  }  & {$-1$} & {$-1$} & {  1} \cr
 {  1}  & {  }  & {  1} & {  1} & {$-1$} \cr
 {  1}  & {  }  & {$-1$} & {$-1$} & {$-1$} \cr
 {   }  & {  }  & {  } &  { } & {  } \cr
 {  2}  &       & $\pmatrix {1 & 0\cr
                             0 &-1\cr}$ & 
                         {$\pmatrix {-1 & 0\cr
                                      0 & 1\cr}$}&
                                 {$\pmatrix{ 0 & 1 \cr
                                             1 & 0 \cr}$} \cr
 { } &{ }&{ }&{ } & { } \cr
 {  2}  & {$\tau \in (0,1)$} & 
           {$\pmatrix{\tau & \gamma \cr
                      \gamma &-\tau \cr}$} & 
                    {$\pmatrix{\tau    & - {\gamma} \cr
                    -{ \gamma} & -\tau \cr}$} & 
                               {$ \pm \pmatrix {1 & 0 \cr
                                                0 &-1 \cr}$} 
\cr 
 {  }  & { $\gamma ={\sqrt{1-\tau^2}}>0$} & { } & { } & { }\cr
\noalign{\medskip}
\noalign{\hrule}
\noalign{\smallskip}
\noalign{\hrule}
  }}$$}

Since the little subgroup of $S_2$ is trivial in this case, we may
interpret the corresponding UIR as describing non-identical geons.
Hence, the question of statistics does not arise with this type of
interpretation, as discussed earlier in \S\S 4 and 5.   Observe, however,
that in this case the geons are indistinguishable in themselves.  It is
only the manner in which they slide through each other that differentiates
them (specifically the relative minus sign in lines 3 and 4 of Table 1).

\bigskip

For future reference, we note here that it is possible to diagonalize $R(E)$
by a simple rotation.  Doing so brings our representation to the form
\eqn\fref
{
  R(s_1)=\pmatrix{0 & 1 \cr 1 & 0} = - R(s_2), 
\ \ R(E)=\pmatrix{1& { 0}\cr 0& {-1}}
}

\noindent 
{\bf{The orbits $\{ \Gamma_\tau \}$, $0<\tau<1$}}

Once again, the little group $K_0$ of {\lgr} is $S_2=\Z_2$ and $\O$ consists
of a single element $\Gamma=\Gamma_\tau$.  Hence the ``core''
representation {\threee} is again the whole of $R$.  Now however, $\Gamma$ is
2-dimensional and more of the machinery of \S 3 comes into play.
With our choice of 
\eqn\SofE
{
    S(E) = \pmatrix{1 & { 0} \cr 0 & {-1}}
}
the projective multiplier $\sigma$ is trivial (as it must be for $\Z_2$)
and $R$ is determined by a UIR $T$ of $K_0=\Z_2$, or in other words
by a sign:
$$
   T(E) = \pm 1 .
$$
The resulting representation $R$ is then given (since $T$ is a
1-dimensional representation) by 
$$
  R(s_i) = \Gamma(s_i)\otimes\One_1 = \Gamma(s_i) \qquad (i=1,2) 
$$
$$
R(E) = S(E)\otimes T(E) = S(E)T(E) = \pm S(E)
$$
or in our chosen basis
\eqn\xtau
{
  R(E) = \pm \pmatrix{ 1 & { 0} \cr 0 & {-1} }
}
with $R(s_i)$ as in Table 1.  This pair of UIR's provides the last line of
Table 2, thereby completing our construction of ${\hat G}$.

What shall we say about statistics in this case?  From the little group's
being $S_2$ we may conclude that the geons are identical.  Moreover
following \S 4, we would naturally associate the statistics of the geons
with the representation $T$ of $S_2$. That is, we would be tempted to say
that the plus sign in {\xtau} describes bosons and the minus sign fermions.
However, such an identification appears to suffer from an ambiguity not
present in the case of the particle group ${\tilde G}$.  There we could
make a {\it canonical } choice of the UIR $S$, based on the fact that
$\Gamma_0$ was a tensor product of 1-geon UIR's.  Here $\Gamma_\tau$ lacks
such extra structure, and there seems no good reason why we couldn't
equally well reverse the sign in our choice of $S(E)$ in \SofE, thereby
interchanging the designations ``fermionic'' and ``bosonic''.  (It is true
that we can ``tell the difference'' between the two choices, but what seems
lacking is any grounds for singling out either choice of $S(E)$ as yielding
``simple exchange without a minus sign''.)  In the face of this ambiguity,
it may be best just to say that the two signs in the last line of Table 2
express an alternative {\it analogous} to the bose-fermi one, but not
necessarily {\it identifiable} with it.

The confusion is heightened if we ask ourselves how the UIR's of Table 2
correspond to those of Table 1 (after all, both the groups $G$ and $\cS$
are isomorphic to $\Z_2{*}\Z_2$).  For the 1-dimensional representations
the correspondence is easy (just omit the $s_2-$column from Table 2 to
obtain Table 1), but for the 2-dimensional representations the
correspondence is less obvious because there are three families of
representations in Table 2 and only one in Table 1.  To reveal the
correspondence, it suffices to rewrite the representations in Table 2.
First notice that the sign in $R(E)$ can be swapped for a sign in $\tau$,
which then ranges over $(-1,1)\backslash\{0\}$.  The penultimate line in
Table 2 then provides the missing $\tau=0$ case of this family, if we write
it in the form \fref.  Thus the entire set of two-dimensional
representations of Table 2 can be seen to make up an open interval's worth
of UIR's, exactly as in Table 1.  The only difference is that in Table 2
the interval is parameterized by $\tau\in(-1,1)$, instead of by
$\tau\in(0,1)$.  From this point of view, the two-dimensional
representations of $G$ are drawn from a single continuum, and it seems
peculiar that one point of that continuum should describe distinguishable
geons, when all the others describe identical ones.

\noindent 
\underbar{An alternative viewpoint}

In fact there exists an alternative point of view (hinted at in \S 5), from
which the geons are identical in all cases, and it may be that this view is
the more appropriate one in some situations.  Indeed, let us return for a
moment to the $p-n$ system which, we argued earlier, was analogous to a
pair of identical primes carrying inequivalent representations of the
1-geon MCG.  Formally, it would be possible to regard the exchange of $p$
with $n$ as an exchange of identical particles, but then {\it both} types
of statistics would be present simultaneously, depending on the parity of
the center of mass wave function.\footnote{*}
{Here we are not referring to the common practice of regarding proton and
neutron as different isospin states of a single particle, the nucleon.
That would be to embed the $p-n$ state-space $\cH$ in a larger state-space
including the $p-p$ and $n-n$ systems as well as the $p-n$.  Rather we mean
to leave $\cH$ unchanged but pretend that the proton and neutron are
indistinguishable, defining bose and fermi sectors of $\cH$ with respect to
the operator which exchanges the (space {\it and} spin) coordinates of $p$
and $n$.}
Processes which changed this parity would be regarded as changing the
statistics of the particles --- in effect changing their identity.  Such a
point of view would seem to be rather useless for an actual $p-n$ system,
but it might make more sense if it were very difficult in practice to
distinguish a proton from a neutron (thus, if processes mixing our putative
``bosonic'' and ``fermionic'' states occurred on longer time scales than
processes taking place {\it within} these subspaces).  Indeed, a description
very much like this one is common in talking about ortho- and para-hydrogen
as if they were composed of different kinds of constituents.

Now in the case of slides, one can imagine situations (a dilute gas of
geons, for instance, or geon--geon scattering at low energies) where
exchanges of location occur much more frequently than collisions in which
one geon penetrates the other.  In such a situation, each 2-dimensional
representation of table 2 would effectively decouple into a pair of
1-dimensional subrepresentations, corresponding to $R(E)=\pm 1$.  The
$R(E)=+1$ subspace could then be interpreted naturally as describing
identical bosonic geons, the other subspace identical fermionic geons, and
transitions between the two subspaces would be induced (relatively
infrequently) only by special kinds of inelastic collisions (slides).
Within this interpretation, moreover, all the 2-dimensional representations
of Table 2 would be on the same footing, nor would we be using ``nonlocal''
effects (the slides) to determine geon identity.  Depending on the
dynamical situation, such a viewpoint might sometimes be preferable to the
more ``kinematical'' one we have adopted in the present paper.  On the
other hand, the method we have adopted for defining particle identity and
statistics is strongly suggested by the mathematics of \S 3; and, as we
have seen, it convincingly interprets the representations in which the
slides act trivially (\S 4).  Here, we wished only to raise the possibility
of an alternative description.  In the sequel we will stick, for clarity,
to our standard mode of expression, in which particle identity and
statistics are to be deduced from the little subgroup $K_0\subseteq{}S_N$
and its PUIR \ $T$.  The difference in any case is only one of convenience;
it is the representation $R$ as a whole that determines the physics, not
its decomposition into one or another type of sub-representation.

\noindent {\underbar {\bf Three $\R{P}^3$ geons}}

The mapping class group when $N=3$ is $G={\cS}\semidirect{}S_3$.  Now,
$\cS$ is generated by $3\times{2}=6$ slides $s_i^j$ satisfying the
commutation relations quoted in \S 2, which here reduce to
$s_i^j\natural{}s_k^j$ and $s_i^js_k^j\natural s_i^k$.  Although we cannot
offer a full classification in this case, one can certainly find {\it some}
UIR's of $\cS$ and use them to construct UIR's of $G$.  

We list in table 3 the UIR's of $G$ which result from the simplest abelian
UIR's $\Gamma$ of $\cS$.  (Notice that for abelian $\Gamma$, the PUIR $S$
of $K_0$ is trivial; hence there is no phase ambiguity in the UIR $T$ of
$K_0$, and our identification of the geon statistics resulting from a given
choice of $T$ is correspondingly unambiguous.) For example, in (a) we have
taken the trivial representation of $\cS$, and hence the corresponding
little group is $K_0=S_3$; then all the geons are identical and can either
obey fermi, bose or parastatistics.  For all the other UIR's listed in the
table, with the exception of (j), the slides render the geons wholly or
partly distinct; either all three are different ($K_0=\{e\}$) or two are
the same but differ from the third ($K_0=S_2$).
In (j) however, $K_0= {\Z}_3$ implying that in this case statistics is
determined {\it not} by an $S_n$ subgroup of $S_3$, but by
${\Z}_3\subseteq{}S_3$, the subgroup of cyclic permutations.  In words: the
geons are invariant only under cyclic permutations but not under a simple
exchange, they are ``cyclically identical'' but not pairwise identical!
One is reminded of other situations in physics in which statistics is
determined by a non-permutation group, for example the braid group in the
case of identical particles on a plane.

\bigskip
\vbox{
\centerline {\bf Table 3: Some UIR's for three ${\R}P^3$ geons, 
                 from abelian UIR's of the slides}
\centerline {In the first column, all unspecified $\Gamma_0({s_i}^j)=1$ }

\baselineskip=12pt
\hsize=7 truein 

$$
\halign{# \hfil & \quad 
        # \hfil & \quad 
        # \hfil & \quad 
        # \hfil & \quad 
        # \hfil & \quad 
        # \hfil & \quad 
        # \hfil & \quad # \cr
\noalign{\hrule}
\noalign{\smallskip}
\noalign{\hrule}
\noalign{\medskip}
{}              & 
Fiducial        & 
Little          & 
$\nu=$          &
UIR   $T$       &
${\tR}=$        &
Identity: 
\cr 
{}                 & 
UIR of ${\cal S}$  & 
group              & 
$|S_3/K_0|$        & 
of $K_0$           &
$(\bigoplus_{i=0}^{\nu-1}\Gamma_i) \otimes {\bf 1}_\beta$  & 
Statistics 
\cr
     {}  &  $\Gamma_0$  &  $K_0$  &  $=|\O|$  &  {}  &  {}  &  {} 
\cr
\noalign{\medskip}
\noalign{\hrule}
\noalign{\medskip}
(a)             & 
 trivial        &
 $S_3$          & 
 1 		& 
 2 1-d 		& 
 ${\bf 1}_{\beta}$	& 
 all identical: 
\cr 
     &    &    &    & 1 2-d &  & Fermi, Bose,
\cr
     &    &    &    &       &  & para 
\cr
     & & & & & &  
\cr
(b) & ${\Gamma_0}({s_3}^2)=-1$ & $\{e\}$ & 6 &
    trivial & $\bigoplus_{i=0}^{5}{\Gamma}_i $ & all distinct
\cr
  {} & {}  &{}&{} &{}& {}& 
\cr
(c) & ${\Gamma}_0({s_3}^1),$& $\{e\}$&6 &
  trivial&$\bigoplus_{i=0}^{5}{\Gamma}_i $ & all distinct 
\cr
 {} & ${\Gamma}_0({s_1}^{2})=-1$ & {}&{} &{}& {}& 
\cr
 {}& { } &{}&{}&{}&{}&{}
\cr
(d) & ${\Gamma}_0({s_3}^2),$& $\Z_2$&3 & 2 1-d &
    $\bigoplus_{i=0}^{2}{\Gamma}_i$ & 2 identical: 
\cr
 {} & ${\Gamma}_0({s_2}^{3})=-1$ &{}&{} &{}& {}& Fermi, Bose  
\cr
 {}&  { }  &{}&{}&{}&{}& {}
\cr
(e) & ${\Gamma}_0({s_1}^2),$& $\Z_2$&3 &2
1-d&$\bigoplus_{i=0}^{2}{\Gamma}_i$& 2 identical: 
\cr
{} & ${\Gamma}_0({s_1}^{3})=-1$ &{}&{} &{}& {}& Fermi, Bose  
\cr
{}& { } &{}&{}&{}&{}& {}
\cr
(f) & ${\Gamma}_0({s_1}^2),$& $\Z_2$&3 &2
1-d&$\bigoplus_{i=0}^{2}{\Gamma}_i$& 2 identical: 
\cr
{} & ${\Gamma}_0({s_3}^{2})=-1$ &{}&{} &{}& {}& Fermi, Bose  
\cr
{}& { } &{}&{}&{}&{}& {}
\cr
(g) & ${\Gamma}_0({s_1}^2),$& $\{e\}$&6
&trivial&$\bigoplus_{i=0}^{5}{\Gamma}_i$& all distinct 
\cr
{} & ${\Gamma}_0({s_1}^{3}),$&{}&{} &{}& {}& 
\cr
{}&${\Gamma}_0({s_2}^1)=-1$ &{}&{}&{}&{}& {}
\cr
{}& { } &{}&{}&{}&{}& {}
\cr
(h) & ${\Gamma}_0({s_1}^2),$& $\{e\}$&6
&trivial&$\bigoplus_{i=0}^{5}{\Gamma}_i$& all distinct 
\cr
{} & ${\Gamma}_0({s_3}^{2}),$&{}&{} &{}& {}& 
\cr
{}&${\Gamma}_0({s_3}^1)=-1$ &{}&{}&{}&{}& {}
\cr
{}& { } &{}&{}&{}&{}& {}
\cr
(i) & ${\Gamma}_0({s_1}^2),$& $\{e\}$&6
&trivial&$\bigoplus_{i=0}^{5}{\Gamma}_i$& all distinct 
\cr
{} & ${\Gamma}_0({s_2}^{1}),$&{}&{} &{}& {}& 
\cr
{}&${\Gamma}_0({s_3}^1)=-1$ &{}&{}&{}&{}& {}
\cr
{}& { } &{}&{}&{}&{}& {}
\cr
(j)  & ${\Gamma}_0({s_1}^2),$& $\Z_3$&2 &3
 1-d & $\bigoplus_{i=0}^{1}{\Gamma}_i$ & ``cyclically identical'': 
\cr
  {} & ${\Gamma}_0({s_2}^{3}),$ & {} & {} & {} & {} & 3 types
\cr
  {}&${\Gamma}_0({s_3}^1)=-1$ &{}&{}&{}&{}& {}
\cr
{}& { } &{}&{}&{}&{}& {}
\cr
\noalign{\medskip}
\noalign{\hrule}
\noalign{\smallskip}
\noalign{\hrule}
}$$}

It is relatively easy to see from this example that as the number of
$\R{P}^3$ primes increases, there will be an increasing number of sectors
in which the statistics is dictated by non permutation subgroups of $S_N$.
For example, it is readily seen that any cyclic subgroup of $S_N$ can play
this role.  In fact, given that every finite group is isomorphic to a
subgroup of a permutation group, it seems plausible that statistics can
be determined by an arbitrary finite group, given a large enough number of
$\R{P}^3$ primes.


\vskip 0.4 true in 
\centerline {\bf \S 7. Conclusions}
\Nobreak
\bigskip

Our main goal in this paper has been to understand physically the various
conceivable quantum sectors (``$\theta-$sectors'') which arise in
4-dimensional quantum gravity when the spatial topology is non-trivial, but
also non-dynamical.  Mathematically, the task of finding all the quantum
sectors translates (if we ignore the possibility of ``Wess-Zumino terms''
or other topological contributions to the quantum amplitude which modify
the classical limit of the theory) into the problem of classifying the
UIR's of the mapping class group $G$ of the spatial 3-manifold $\M$.
Ideally one would wish for a full classification, but that is out of reach
at present, if only because the prime 3-manifolds themselves remain
unclassified.  In the present paper we have concentrated on bringing out
the specific features which arise because more than one geon is present.

In general $\M$ will be a connected sum of prime manifolds, and 
we have based our physical interpretation on a picture which
regards each such prime (excluding the handles) as giving rise, on
quantization, to a physical particle or {\it topological geon}.  Our
analysis has aimed at characterizing the UIR's of $G$ in the particle
language appropriate to such a geon interpretation.

In one important case, our analysis has been complete in the sense that we
have classified all possible multi-geon UIR's in terms of the UIR's of the
mapping class groups, $G_i$, of the individual primes.  The case we are
referring to is the one in which the slide-diffeomorphisms are represented
trivially, so that $G$ reduces in effect to the ``particle subgroup''
${\tilde G}$ generated by the internal diffeomorphisms of the individual
primes and the permutations of identical primes among themselves.  Our
classification of the UIR's of $\tG$ was based on the decomposition of
$\tG$ as a semidirect product of the internal diffeomorphisms $G_{int}$
with the permutations $S_N$.  From this decomposition there followed a
characterization of the UIR's of $G$ in terms of the UIR's of the $G_{i}$
on one hand, and the UIR's of certain subgroups (``little groups'') of the
permutations on the other hand.  (The mathematical facts used in obtaining
the classification were summarized in \S 3 above.)

The characterization of the UIR's of $G$ in this manner lends itself to a
mode of expression in which the ``internal'' UIR's (i.e. the UIR's of the
internal diffeo groups $G_i$) determine the physical identity of the
corresponding geons (their ``quantum numbers'' or ``species''), while the
UIR of the resulting little subgroup of $S_N$ determines the {\it
statistics} of the geons.  In this language, the classification scheme of
\S 4  states that one obtains the general UIR of $\tG$ by ($i$)
choosing a geon species for each prime which is present (i.e. a UIR of its
MCG), and then ($ii$) choosing a statistics for each species of geon
(i.e. a UIR of $S_n$, where $n$ is the number of geons of that species
which are present).  Notice that in general there will be more types of
geons present than distinct diffeomorphism classes of primes, because a
single type of prime can carry many different UIR's of its internal
diffeomorphism group: ``quantum breaking of indistinguishability''.  This
mode of description of the UIR's seems fully satisfactory from a physical
point of view, and its success in turn should offer strong support to the
interpretation of primes as particles.

In the more general case of UIR's $R$ of $G$ for which the slides are not
trivial, one may attempt a similar analysis based on the fact that the full
mapping class group $G$ admits a (two--stage) decomposition into semidirect
products.  Once again, this analysis suggests definite criteria for when
the geons should be called identical, and for what type of statistics the
identical geons then exhibit.  When slides are involved however, the
harmony between the group theoretical analysis and the physical behavior of
the geons is less perfect than for $\tG$ alone, the problem deriving in
some sense from the nonlocal character of the slides.  Indeed we have
sketched in \S 6 an alternative mode of description which would interpret
geon identity and statistics somewhat differently, and it should be borne
in mind that some of the detailed points we will make in the summary below
would have to be stated differently if the alternative mode of description
were adopted.

Aside from this purely interpretational difficulty resulting from the
slides, a more serious problem is that the general form of the UIR's of the
slide subgroup $\cS$ is not known, and hence a general account of the UIR's
of $G$ cannot be given.  However, one may still start with some UIR's
${\Gamma}_0$ of $\cS$ that one does know (e.g., abelian UIR's) and use them
to build up a family of UIR's of $G$.  In this manner, a large number of
``theta sectors'' can be found, and some of them introduce new effects,
including previously unknown possibilities for particle identity and
statistics, as described earlier in \S\S 5 and 6.

By way of summary, we now list some of the general conclusions on geon
identity, statistics, and ``internal symmetry'' which have emerged
from the analysis of this paper.

\noindent (a) {\bf  Geon identity}
\Nobreak
\item{(i)} 
Primes that are diffeomorphic, and therefore classically identical, will
give rise to distinguishable geons if their internal groups are represented
inequivalently.  (For example, if $\cal P$ is a prime with
${\cal{R}}_{2\pi}\neq e$ and if $\M = \R \# {\cal P} \# {\cal P}$, then
representing one of the ${\cal R}_{2\pi}$'s by $+1$ and the other by $-1$
will result in two geons of opposite spin-type, one tensorial and the other
spinorial.)

\item{(ii)}
Even if two primes carry the same representation in the sense of (i), it is
possible for the slides to 
render them distinguishable.  Examining the internal structures of
individual geons is therefore inadequate to determine whether they are
truly identical; one must study  geon-geon collisions as well.

\noindent (b) {\bf Geon Statistics }
\Nobreak
\item{(i)} 
There will always exist sectors in which the spin-statistics correlation
for geons is violated.  Indeed, this occurs for an arbitrary prime $\cP$
and an arbitrary geon type based on $\cP$ (i.e. an arbitrary UIR of the MCG
of $\cP\#\R^3$).  In particular, the geons can be arranged to be all
bosonic or all fermionic irrespective of whether they are tensorial or
spinorial.  (How the internal diffeomorphism group of a prime is
represented determines whether the corresponding geon is spinorial or
tensorial.)

\item {(ii)} 
When $N$ $\geq$ 3 there will always exist sectors in which the geons obey
parastatistics.

\item {(iii)} 
In the case of $\R{P}^3$ geons, when $N\geq 3$ there will always exist
sectors in which geon identity and statistics are expressed not by a
permutation subgroup of $S_N$ but by (for example) a cyclic subgroup of
$S_N$.  (We constructed such sectors in \S 6 using certain abelian
representations of the slide subgroup). This type of behavior seems to be
new.  Moreover, Balachandran has pointed out that the same thing can in
principle occur for string-like excitations in suitable condensed matter
systems {\BalPrivate}.

\item{(iv)}
For $N\ge 4$ it  may be that there also exist sectors in which statistics
is expressed by 
a {\it projective} representation of $S_N$ or its subgroups, rather than an
ordinary representation.

\item{(v)}
For a pair of $\R{P}^3$ geons, there exist quantum sectors for which the
bose-fermi distinction becomes ambiguous: one has pairs of sectors differing
only by the sign of the exchange, without being able to say which sector is
the bosonic one and which the fermionic.

\noindent (c) {\bf Internal Symmetry of Geons}
\par\nobreak

\item{(i)}
There exist quantum sectors for certain three manifolds in which the slides
render the geons internally asymmetric even though classically the primes
possess symmetries (expressed by their internal diffeo groups).  Thus the
examination of an isolated prime $\cal P$ is not adequate to determine its
quantum symmetries; one must look at geon-geon collisions as well.

The existence of such novel aspects of geon behavior as listed above
bespeaks a richness of quantum gravity deriving from the possibility of
non-Euclidean {\it topology} inherent in general relativity.  In itself,
this novelty is intriguing, but not at all disturbing.  However, what would
appear to be disturbing is the huge degree of ambiguity in quantization
associated with all these different UIR's of $G$.  For example, even in the
very simplest case where $\M={\R}^3\#\R{P}^3\#\R{P}^3$, we saw that there
arise an infinite number of two dimensional sectors parameterized by a real
number $\tau$.  Within the particle picture this implies an infinite number
of possible geon-types for nature to choose from (even though the spatial
topology is fixed).  And with more (and more generic) primes present, one
would expect far more free parameters, if not a discrete infinity of
possible sectors.\footnote{*}%
{It is not clear whether the free parameters which arise in this manner
from spatial topology are related to the free parameters proposed to arise
in connection with certain types of topological {\it fluctuations}
(``wormholes'') {\wormhole}.}
This does not seem to accord with the conception of quantum gravity as a
fundamental theory.  (Indeed, even classically, it seems a mystery how
nature could have chosen one out of the discrete infinity of possible
{\it topologies} for $\M$.)

\bflb~One might object at this point that perhaps our last complaint is not
well founded because not all these sectors are really possible dynamically.
Indeed, our analysis so far has been on a purely ``kinematical'' level, in
the sense that, basically, we have used nothing more than the facts that
general relativity is a generally covariant theory possessed of a metric
field $g_{ab}$.  We have not used the specific dynamics of the
Einstein-Hilbert Lagrangian (manifested in a canonical setting by the
Hamiltonian constraints), so how do we know that this dynamics will not
exclude most of the UIR's we have been examining?  In view of this possible
objection, we will pause for a moment and try to make it plausible that in
fact no sector is excluded by the dynamics.  For definiteness we conduct
the discussion in the language of canonical quantization rather than that
of the path-integral.

\bflb~According to reference {\dwitt}, there is in 3-dimensions no topological
obstruction to the specification of asymptotically flat initial data.  Thus
for each $\M$ we have considered, there exists at least one classical
3--metric $^3g_{ab}$ which (together with an appropriate extrinsic
curvature) solves the classical initial value constraints.  Now classical
general relativity is by definition the classical limit of quantum gravity;
moreover we may scale our classical initial data to be as far as desired
from the Planck scale.  Hence we may reasonably assume the existence in the
quantum theory of a WKB wavefunction $\psi$ peaked at a single 3-metric
$^3g$, or rather on the set of $[^3g]$'s obtained by the different allowed
slicings of the classical spacetime corresponding to our initial data,
where $[^3g]$ denotes here the set of metrics related to $^3g$ by {\it
small} diffeomorphisms.  (Recall that invariance under the small
diffeomorphism group $\Diff_0^\infty(\M)$ suffices to satisfy the momentum
constraints.)

\bflb~Under the action of a nontrivial element of the mapping class group $G$,
the metric $^3g$ goes into a gauge-related metric $^3g'$, which, by
definition belongs to a {\it different} equivalence class from that of
$^3g$, i.e.  $[^3g]\not=[^3g']$.  Thus, $G$ takes the equivalence class
[$^3g$] into an arbitrary element of the fiber of
${\cal{R}}/\Diff_0^{\infty}$ which lies over the 3-geometry
$q=cls(^3g)\in{}Q={\cal{R}}/\Diff^{\infty}$.  The fact that the images of
$[^3g]$ by the elements of $G$ are all distinct elements of the fiber
(remember that the action of $\Diff^\infty$ is free) makes it plausible
that the images of $\psi$ by $G$ are all linearly independent.  (In general
a given 3-geometry will occur only once among the slices of 4-metric.  Note
also, that we don't really need independence of {\it all} the images when
this number is infinite; if we only want to reproduce a finite dimensional
UIR, it will suffice to choose $^3g$ ``macroscopic enough'' so that we have
a sufficiently large finite independent set.)

\bflb~The set of WKB wavefunctions $\{{\psi}_i \}$ obtained in this way spans a
vector subspace $\cH$ of the full Hilbert space which naturally carries the
regular representation of $G$ (or some sufficiently large portion thereof).
Hence any UIR will be realizable in $\cH$, whence in the state space of our
theory.  (Since $\psi$ satisfies all the constraints, so will the elements
of $\cH$.)  By this reasoning, our analysis of the possible quantum sectors
should be independent of the particular dynamics imposed, as long as that
dynamics allows arbitrary topologies to occur in the classical limit.~\bfrb

As we have alluded to more than once, the plethora of quantum sectors
described in this paper arises in theories of quantum gravity in which the
spatial topology is ``frozen''.  A natural question to ask is whether
``thawing'' the topology would remove the redundant sectors, among which
the most plainly unphysical are those which violate the spin-statistics
correlation.  The most obvious framework in which the spacetime topology
can fluctuate is the sum-over-histories (cf.  {\ArvindRafael}), and a set
of rules for regaining the spin-statistics correlation in such a framework
has been suggested in {\amalfi}.  In {\rdsfay} it has been shown that, even
without imposing any new conditions like those of {\amalfi}, the
sum-over-histories automatically reproduces the spin-statistics correlation
for certain pairs of identical geons (lens spaces) formed via a certain
type of ``U-tube cobordism''.  Other examples where topology change
excludes sectors of the frozen theory have been given in {\jimdon} and
{\jorma}.

The existence of redundant quantum sectors is not limited to quantum
gravity.  An example from molecular physics is the ``rigid Born-Oppenheimer
approximation''.  There the electronic degrees of freedom and the
fluctuations in the shape of the molecule have been ``frozen'', and one is
left with a rigid 3-dimensional body.  When the shape of the body is
symmetrical, the resulting configuration space is multiply connected, and
this leads to inequivalent quantum sectors in the familiar manner
(including sectors in which a kind of ``quantum chirality'' or ``parity
breaking'' occurs {\balwitt}).  However, not all of these sectors are truly
possible for a given molecule.  Rather the rigid body is only an effective
description of the molecule.  When one restores the degrees of freedom
underlying this effective description (its ``material base''), one finds
{\balsachin} that some of the sectors are removed and the others brought
together into a unified state space (unified because the strict
superselection rule separating the sectors is lifted.)

We may take this as a typical pattern and a clue to what is happening in
quantum gravity.  On this view, incorporating topology change should
restore some of the missing degrees of freedom and remove some of the
redundancy.  We suspect however that topological fluctuations alone will
not suffice to remove the infinity of quantum sectors, and that the
remaining redundancies will disappear only in a deeper theory based on a
discrete structure like that of the causal set {\posets}.

\bigskip
\centerline {\bf Acknowledgments}
\Nobreak
We would like to thank Fay Dowker for several discussions.  We also would
like to thank A.P. Balachandran for reading the first draft of this paper
and for the many discussions that followed.  Finally, we are grateful to
Nico Giulini for a key suggestion in connection with the Appendix, and for
providing us with related references.  This research was partly supported
by NSF grant PHY-9600620 and by a grant from the Office of Research and
Computing of Syracuse University.

\vskip 0.4 true in 
\centerline {\bf Appendix}
\Nobreak
\bigskip

\def\Nbar{{\overline N}}
\def\Kbar{{\overline K}}
\def\Gbar{{\overline G}}
\def\slidebar{{\overline\slides}}
\def\internalbar{{\overline\internals}}
\def\permbar{{\overline\perms}}
\def\partgrp{{\widetilde G}}		

In the main text we described generators for the mapping class group $G$
which go by the names of ``slides'', ``internal diffeos'' and
``exchanges'', and we designated the three subgroups of $G$ which these
generate by the respective symbols $\slides$, $\internals$ and $\perms$.
We also demonstrated, using developments, that $\slides$ is a normal
subgroup of $G$ and that $\internals$ is invariant under conjugation by
elements of $\perms$.  This suffices to establish the inclusions of normal
subgroups,
$\slides\subseteq\slides\internals\subseteq\slides\internals\perms=G$ and
$\internals\subseteq\internals\perms=\tG$.  If we knew in addition that 
$(\slides\internals)/\slides=\internals$,
$G/\slides\internals=\perms$, and 
$G/\slides=\tG$, 
then we would have established the semidirect product decomposition {\comp}
of \S 2 on which we based most of our analysis.  In this Appendix,
assembling most of the necessary ingredients from the literature, we sketch
a proof of the above equalities, or rather a proof of {\comp} itself,
together with certain other assertions we made in \S 2.
Specifically we demonstrate the following facts.  (Our notation for
semidirect product has been explained in \S 2.)

\noindent {\bf Facts to be shown}

\item{(0)} $G = \slides \semidirect {\internals} \semidirect \perms$ 
\item{(1)} $\perms$ is isomorphic to the permutation group on $N$ elements
             (as its name implies).
\item{(2)} $G_{int} = G_1 \times G_2 \times \cdots \times G_N $
\item{(3)} $\slides$ is described correctly by the generators and relations
                     given in Section 2.

\noindent {\bf Lemma}
Let $G=NK$ with $N$ normal in $G$, let $\Gbar=\Nbar\semidirect\Kbar$, and
     let $f:G\to\Gbar$ such that
  \item{(i)} $f(N)\subseteq\Nbar$ and $f(K)\subseteq\Kbar$
  \item{(ii)} either 
    {(a)}  $f|N$ is injective, or
    {(b)}  $f|K$ is injective
  \item{(iii)} $f$ is surjective.

  \item{} Then $ G = N \semidirect K. $

  \item{} Moreover in
     {case (a)}  $N \isom\Nbar$, and in
     {case (b)}  $K \isom\Kbar$.
 
\noindent Proof: straightforward.

Henceforth we will assume the truth of the Poincar\'e conjecture.  We need
this, because we will rely on the standard device of replacing
diffeomorphisms by their action on loops in $\M$, and such a device
obviously would not succeed in connection with primes which have no
nontrivial loops.

In the following we write $\pi$ for $\pi_1(\M)$, $\Phi$ for the
homomorphism of $G$ into $\Aut(\pi)$ which takes a diffeomorphism into its
natural action on loops in $\M$, and $\Gbar$ for $\Phi(G)$, the image of
$G$ in $\Aut(\pi)$.  Similarly, we write $\slidebar$, $\internalbar$ and
$\permbar$ for the images of these subgroups in $\Aut(\pi)$.

\noindent 
{\bf Lemma on the structure of $\Gbar$ }
$$  
     \Gbar = \slidebar \semidirect\internalbar \semidirect\permbar    
$$
If we use the generators described in the text, then this lemma follows
directly from the structure of the Fuks-Rabinovitch (F-R) relations, as
presented in {\fuksrab \mcmil}.  In fact all of the latter are relations
{\it{}within} one of the three subgroups of the lemma, or else they specify
how one of the subgroups {\it acts on} the generators of another.  This,
plus the obvious fact that the exchanges leave $\internalbar$ invariant, is
all we need to establish the lemma.  (In order to be sure that the F-R
relations are complete, we must know that the individual factors
$\pi_1(\cP_j)$ of $\pi$ are not further decomposable as free products.  This
is the content of ``Kneser's Conjecture'', Theorem 7.1 of {\hempel}.)

We take from {\mcc} the following

\noindent 
{\bf Fact}  $\ker(\Phi)\subseteq\internals$.

(In comparing with that reference, notice that ``rotations parallel to the
connected sum 2-spheres'' are included in what we call $\internals$;
physically such a diffeomorphism represents a $2\pi$ rotation of the
corresponding geon.)  Further, we have,

\noindent 
{\bf Claim 1} $\Phi|\perms$ is injective 

   This follows trivially from fact that $\pi=\pi_1(\M)$ is the free
   product of the individual $\pi_1(\cP_i)$ (all of which are nontrivial,
   by the Poincar\'e conjecture): for any permutation $p$ we can find an
   element of one $\pi_1$ which goes to an element of a different $\pi_1$,
   and these cannot be equal in a free product.

\noindent 
{\bf Claim 2} $\Phi|\slides$ is injective 

  This follows from the fact that the generators of $\slides$ (i.e. the
  slide diffeomorphisms described in Section 2) already satisfy all of the
  F-R relations that involve only the ``slide generators'' of $\Gbar$,
  which in turn can be seen, for example, by using developments as in \S 2.

Now let us apply our first lemma with $N=\slides$, $K={\internals}\perms$,
$G=G$ and $f=\Phi$.  We know that $N$ is normal in $G$, while
$\Gbar=\Nbar\semidirect\Kbar$ follows from our second lemma.  It is trivial
that $f$ is surjective and that $f(N)\subseteq\Nbar$, $f(K)\subseteq\Kbar$.
Finally we have that $f|N$ is injective from Claim 2 above.  Hence case (a)
of the lemma applies and tells us both that
$$
        G = \slides \semidirect (\internals\perms)
$$
and that $\slides = \slidebar$.  The latter is item (3) in our list of
facts to be proven, because the commutation relations quoted in \S 2 are a
complete set of relations for $\slidebar$ according to {\fuksrab \mcmil}.

Now let us apply our first lemma again with $N={\internals}$, $K=\perms$
and with $G$ being the ``particle group'', ${\internals}\perms=\tG$.  By
Claim 1 above, we are in case (b) of the lemma , so we conclude
$$   
   \partgrp = \internals\semidirect\perms
$$
and also $\perms=\permbar$.  The latter establishes item (1) of our list,
since the generators and relations of $\permbar$ are precisely those of the
permutation group on $N$ elements. 

Putting the last two results together yields
$G=\slides\semidirect(\internals\semidirect\perms)$, which is equivalent to
item (0) in our list of facts to be proven.  

It remains to demonstrate item (2), or in other words equation {\gint} of
the main text.  As mentioned in \S 2, this isomorphism almost follows
trivially from the fact that the each $G_i$ can be defined with support on
its own prime, but because the supports {\it need} not be disjoint, we
strictly speaking have not yet excluded that some combination of elements
of different $G_i$'s might vanish.  In that case, equation {\gint} would
not be an equality, but only a surjective map
$   G_1 \times G_2 \times \cdots \times G_N \to G_{int}. $

The needed proof that this map is in fact injective has been given in 
{\nicoHPA}, using the following argument.  According to theorem 1 of
{\hrksmc},  $\Diff^\infty(\M)$ is a trivial principal fibre bundle with
fibre $D_1\times{}D_2\times\cdots{}D_N$, where 
$D_i=\Diff^\infty(\R^3\#\cP_i)$ is the group of internal diffeomorphisms of
the $i^{th}$ prime.  The base space $Imb$ of this bundle is a space of
embeddings of $B$ into $\M$, where $B$ is the submanifold of $\M$ that
remains when the primes are ``cut out'' of $\M$.  (Specifically, an
embedding belongs to $Imb$ iff it extends to an element of 
$\Diff^\infty(\M)$.)  Thus we have the homeomorphism, 
$$
  \Diff^\infty(\M) \isom D_1 \times D_2 \times \cdots \times D_N \times {Imb}
$$
Since in general $\pi_0(A\times{B})=\pi_0(A)\times\pi_0(B)$, this implies
that the fibre inclusion, 
$$
      D_1 \times D_2 \times \cdots \times D_N \to   \Diff^\infty(\M)
$$
induces an injective homomorphism
$$
\eqalign{
  G_1\times G_2\times\cdots\times G_N 
   & = \ \pi_0(D_1)\times\pi_0(D_2)\times\cdots\times\pi_0(D_N) \cr
   & = \ \pi_0(D_1\times D_2\times\cdots\times D_N) 
      \to \pi_0(\Diff^\infty(\M)) = G .  }
$$
Moreover, the image of $G_1\times G_2\times\cdots\times G_N$ by this
mapping is by definition the internal group $G_{int}$, which therefore is
isomorphic to $G_1\times G_2\times\cdots\times G_N$, as claimed.

\listrefs

\end